\documentclass[%
 aip,
 amsmath,amssymb,
 reprint,%
]{revtex4-1}

\usepackage{graphicx}% Include figure files
\usepackage{dcolumn}% Align table columns on decimal point
\usepackage{bm}% bold math
\usepackage{float}
\usepackage[utf8]{inputenc}
\usepackage[T1]{fontenc}
\usepackage{mathptmx}
\DeclareUnicodeCharacter{2212}{-}

\newcommand{\dGwol}{\Delta G_{\textup{W}\rightarrow\textup{Ol}}}

\begin{document}

\preprint{AIP/123-QED}

\title[]{Resolution limit of data-driven coarse-grained models spanning chemical space}
% Force line breaks with \\

\author{Kiran H. Kanekal}
 \email{kanekal@mpip-mainz.mpg.de}
\affiliation{ Max Planck Institute for Polymer Research, Ackermannweg 10, 55128 Mainz, Germany
}%

\author{Tristan Bereau}
\affiliation{ Max Planck Institute for Polymer Research, Ackermannweg 10, 55128 Mainz, Germany
}%

\date{\today}% It is always \today, today,

\begin{abstract}
Increasing the efficiency of materials design and discovery remains a significant challenge, especially given the prohibitively large size of chemical compound space. The use of a chemically transferable coarse-grained model enables different molecular fragments to map to the same bead type, while also reducing computational expense. These properties further increase screening efficiency, as many compounds are screened through the use of a single coarse-grained simulation, effectively reducing the size of chemical compound space. Here, we propose new criteria for the rational design of coarse-grained models that allows for the optimization of their chemical transferability and evaluate the Martini model within this framework. We further investigate the scope of this chemical transferability by parameterizing three Martini-like models, in which the number of bead types ranges from five to sixteen for the different force fields. We then implement a Bayesian approach to determining which chemical groups are more likely to be present on fragments corresponding to specific bead types for each model. We demonstrate that a level of performance and accuracy comparable to Martini can be obtained by using a force field with fewer bead types. However, the advantage of including more bead types is a reduction of uncertainty with respect to back-mapping these bead types to specific chemistries. Just as reducing the size of the coarse-grained particles leads to a finer mapping of conformational space, increasing the number of bead types yields a finer mapping of chemical compound space. Finally, we note that, due to the relatively large size of the chemical fragments that map to a single martini bead, a clear resolution limit arises when using $\dGwol$ as the only descriptor when coarse-graining chemical compound space.   
\end{abstract}

\maketitle

\section{Introduction}

Molecular design is a cornerstone of materials science, requiring a fundamental understanding of the relationships between molecular structure and the resulting properties. Traditionally, these structure-property relationships\cite{Roy2015} only arise after multiple rounds of screening and discovery of new materials.\cite{Xi2018, Patel2016,mounet2018two,greenaway2018high,burns2018high} These cases are examples of direct molecular design, in which the space of all chemical compounds, known as the chemical compound space (CCS), is explored to determine the most suitable chemistry for the target application. Direct molecular design can be interpreted as identifying a hypersurface in the high-dimensional CCS onto a lower dimensional space defined by certain key molecular descriptors that strongly correlate with the desired property. In contrast, inverse molecular design, in which a structure-property relationship is used to infer a suitable chemical structure from a desired property, remains the holy grail of materials science. The main obstacle to achieving this goal is the inability to quickly establish structure-property relationships that can span broad regions of CCS. This is an exceedingly difficult task, given that the size of CCS was estimated to be 10\textsuperscript{60} for drug-like molecules less than 500 Da.\cite{Dobson2004} Experimentally, this process is inhibited due to both the material and time cost associated with synthesizing and testing a large variety of chemistries that are necessary to infer a relation that is both robust and accurate enough to enable inverse molecular design. 

\begin{figure}[H]
  \begin{center}
    \includegraphics[width=\linewidth]{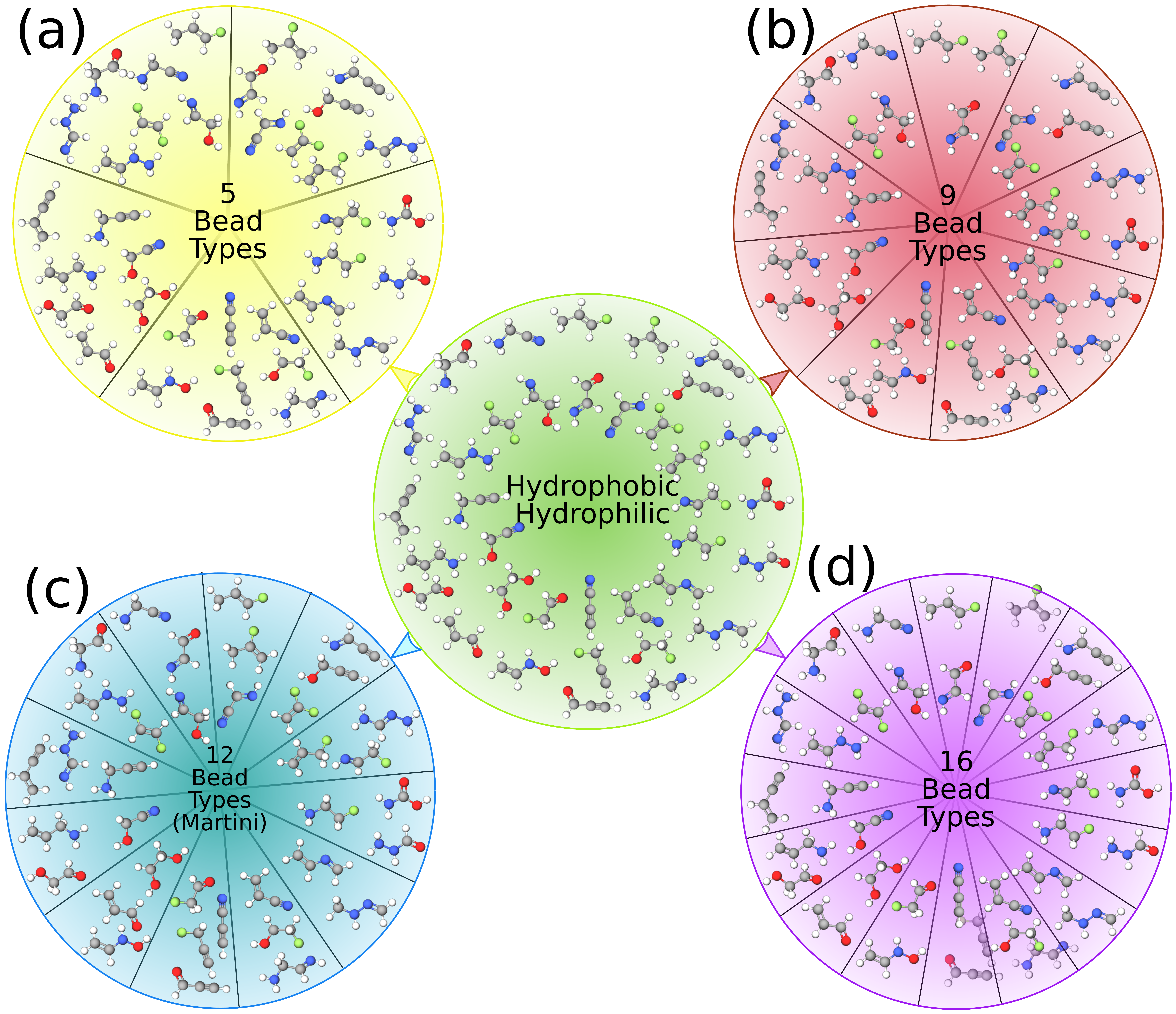}
    \caption{A cartoon schematic showing the projection of CCS onto the hydrophobicity descriptor $\dGwol$, allowing for the creation of top-down chemically-transferable coarse-grained models with a) five, b) nine, c) twelve, and d) sixteen bead types. The number of bead types included in these models defines the degree to which CCS is partitioned on the $\dGwol$ axis. By varying the number of bead types in each model, we obtain greater insight as to the range of chemistries spanned by a single bead type.}
    \label{fig:Intro}
  \end{center}
\end{figure}

Computationally, recent advancements in processing power and in machine learning have enabled several efficient methods for estimating the electronic properties of a large variety of materials.\cite{Faber2018,Willatt2018,ghiringhelli2015big,ramakrishnan2015big,butler2018machine,bereau2018non} These methods have the added benefit of screening molecules that cannot be easily synthesized, and can thus motivate (or demotivate) the experimental exploration of these chemistries. However, there has been relatively little success in applying computational high-throughput screening methods to determine the stability of chemical compounds in soft matter systems for which thermal fluctuations play a critical role.\cite{Bereau2016,bereau2018data} Force field based methods, such as molecular dynamics simulations, are typically used to account for the immense number of configurations that result from thermal fluctuations in these systems. Unfortunately, due to the extensive computational resources required, a high-throughput scheme based on atomistic molecular dynamics simulations is currently unfeasible for spanning the large regions of CCS needed to obtain broadly applicable structure-property relationships. 

Coarse-grained molecular dynamics simulations provide a means to significantly reduce the computational expense relative to fully atomistic simulations while still capturing the relevant physical properties.\cite{Noid2013,kmiecik2016coarse,voth2008coarse,peter2009multiscale} Coarse-grained representations of molecules result from mapping groups of atoms to coarse-grained “pseudo-atoms” or beads. The governing interactions between beads are determined such that the desired properties of the atomistic system are retained. This usually corresponds to a smoothing of the underlying free-energy landscape, allowing for more efficient sampling. Conventionally, coarse-graining is applied to a single molecule with the goal of efficiently sampling a specific system of interest. The coarse-grained potentials are obtained via one of several possible methods (e.g., iterative Boltzmann inversion\cite{schommers1983pair,tschop1998simulation}, force-matching\cite{Ercolessi1994,izvekov2005multiscale}). However these methods are computationally expensive, requiring an initial atomistic simulation that sufficiently explores the underlying free energy landscape of the system of interest.\cite{ayton2007multiscale} Therefore, adapting coarse-grained molecular dynamics simulations to high-throughput screening of chemical compounds requires flexible yet reliable mapping and force field parameterization methods that do not rely on results from higher resolution simulations for each compound screened.

The coarse-grained Martini force field has become widely used to simulate biological systems as it provides a robust set of transferable force field parameters by constructing biomolecules from a small set of bead types.\cite{M-2004,M-2008,Marrink2013} The Martini model is a top-down model, which maps an atomistic compound or molecular fragment to a coarse-grained site based on its partitioning between aqueous and hydrophobic environments. In the context of molecular design, the main advantage that Martini provides is its chemical transferability. While the force field was explicitly parameterized for a set of specific molecules, a single Martini bead can represent several different chemistries that share similar oil/water partitioning characteristics. In this context, the main feature captured by the Martini model is hydrophobicity, which can act as a key driving force in the physics of soft-matter systems. Rather than running a single atomistic simulation that yields a single data point in CCS, a Martini coarse-grained molecular dynamics simulation provides a representative point in CCS, corresponding to the average behavior of all the chemistries that lay in the region surrounding that point. Thus, high-throughput coarse-grained (HTCG) simulations that use chemically-transferable force fields, such as Martini, are advantageous because they span vast regions of CCS to quickly infer the structure-property relationships and chemical descriptors that can be used to enable inverse molecular design at any resolution. Menichetti et al.~ recently demonstrated this by running Martini HTCG simulations to construct a structure-property relationship describing the thermodynamics of the insertion of a small organic molecule into a biological membrane across CCS.\cite{Menichetti2017,Menichetti2019} In doing so, they discovered a linear relationship between the bulk partitioning behavior of the solute and its potential of mean force. They were then able to identify a structure-property hyper surface to obtain membrane permeabilities for these solute molecules. Using the Generated DataBase\cite{Fink2005,fink2007virtual} (GDB), a systematically computer-generated set of organic drug-like compounds, as a proxy for CCS, we then related the regions of this surface to regions of CCS that were dominated by specific chemical moieties, enabling inverse molecular design of small molecules given a desired permeability. The question remains: how representative  of CCS is the Martini force field? Given that Martini was designed to reproduce the partitioning behavior of certain solvents as well as the properties of lipid-bilayer membranes, is there a way to accurately parameterize a transferable coarse-grained force field with the goal of optimizing its coverage of CCS? In the context of high-throughput coarse-grained simulations that use Martini, creating a structure-property relationship that enables inverse design requires an understanding of the chemistry that is representative of a specific bead type. The metric used in assigning specific chemical fragments to Martini bead types is the water/octanol partition free energy ($\dGwol$). Therefore, an intuition for which chemistry maps to a given bead type can only be obtained by understanding how $\dGwol$ varies as a function of chemistry. Given that the number of heavy (non-hydrogen) atoms that usually map to a Martini bead ranges from three to five, we can think of each bead as representing a small carbon scaffold perturbed to some degree by either replacing carbons with other heavy atom types (e.g., oxygen, nitrogen, or fluorine) or by replacing single bonds with double or triple bonds. We define a functional group as being one or a localized combination of these types of perturbations. 

In this work, we quantify the information loss that occurs when a top-down coarse-grained model, like Martini, is used to reduce the resolution of CCS. Additionally, we parameterize three sets of coarse-grained force fields in the Martini framework. In this context, we use the terms ``force field'' and ``model'' interchangeably, defined as a set of parameters which describe the interactions between a fixed number of coarse-grained representations called bead types. Each force field developed in this work consists of five, nine, and sixteen neutral bead types, as well as two extra types to account for hydrogen bond donors and acceptors. We observe that Martini does not provide the most efficient reduction of CCS.  We show that the nine-bead force field reduces CCS to the same degree as Martini despite having three fewer bead types, and that further increasing the number of bead types yields negligible improvements in the performance of the model. The models are validated by performing coarse-grained simulations to calculate the water/octanol partition free energies of approximately 500 compounds for which experimental data is available. Finally, we demonstrate that the main advantage of a force field with a large number of bead types is the reduction of uncertainty when back-mapping these coarse-grained representations to real chemical functional groups. Just as decreasing the resolution of the CG mapping reduces the resolution of the potential energy landscape, a reduction in the number of bead types of a chemically transferable CG force field allows for an increased degeneracy of chemical fragments that map to a single bead type, illustrated in Fig.~\ref{fig:Intro}. Ideally, a well-designed chemically transferable CG force field would contain some number of bead types that can be intuitively back-mapped to single chemical functional groups. However, the size of a single functional group is small relative to the size of a Martini bead, such that many functional groups could be identified within a fragment mapping to a single Martini bead. Here, we demonstrate that this mismatch between the size of a Martini bead and a single functional group requires additional constraints in order to identify the unique chemistry that maps to each bead type. Incorporating these constraints into a Bayesian formalism yields probabilities of specific chemistries mapping to a given bead type, further promoting inverse molecular design. However, even these additional constraints allow for the same functional groups to be present in multiple bead types, indicating a natural resolution limit when using $\dGwol$ as the sole basis for a chemically-transferable, top-down coarse-grained model.

\section*{Methods}

\subsection*{The Auto-Martini Algorithm}
This work relies on the \textsc{auto-martini} algorithm initially developed by Bereau and Kremer.\cite{Bereau2015} The algorithm first determines an optimal mapping for an organic small molecule. The mapping provides the number of coarse-grained beads used to represent the molecule as well as their placement. A mapping cost function is minimized for each molecule so as to optimize both the number and placement of beads used in its coarse-grained representation. The assignment of coarse-grained potentials to each bead (bead-typing) occurs by assigning an existing Martini bead type that has the closest matched water/octanol partition free energy ($\dGwol$) with that of the molecular fragment encapsulated by the bead. The partition coefficients of these fragments are obtained by using \textsc{alogps},\cite{Tetko2001,Tetko2002} a neural network algorithm that predicts these values given the chemical structure of the fragment. In this work, we use an updated version of the \textsc{auto-martini} algorithm that has three significant changes from the previous version. The first change is an increased energetic penalty for ``lonely'' atoms (i.e., atoms that fall outside the Van der Waals radius of the placed coarse-grained beads). The second change is a reduction of the multiplicative factor used when assigning bead types to rings for both five and six-membered rings. Finally, the cutoff value for the $\dGwol$ for the assignment of donor and acceptor fragments to their corresponding bead types was modified such that the CG and atomistic population distributions more closely matched. All of these changes increased the algorithm's accuracy, which is quantified in the supporting information. Using the refined \textsc{auto-martini} algorithm, approximately 3.5 million molecules with ten heavy atoms or less that make up the GDB were mapped to coarse-grained representations for four different force fields. The molecules contain carbon, nitrogen, oxygen, fluorine, and hydrogen atoms only. Of these 3.5 million compounds, approximately 340,000 were successfully mapped to both coarse-grained unimers (1 bead representations) and dimers (2 bead representations) for all of the force fields described in this work. The majority of the remaining compounds were mapped to coarse-grained representations with a higher number of beads, and a small fraction of compounds were unable to be successfully mapped by the algorithm. Histograms comparing the distributions of $\dGwol$ for each set of atomistic compounds mapping to CG unimers and dimers and their CG counterparts were constructed using the \textsc{numpy} histogram function\cite{oliphant2006guide}, with the number of bins equal to 1000 and 1050 for unimers and dimers, respectively. These histograms are shown in Fig.~\ref{fig:Hist_REs}a-d for Martini, while the other histograms can be found in the SI.

\subsection*{The Jensen-Shannon Divergence}
In this work, the  main tool used to quantify information loss when going from atomistic to coarse-grained resolution is the relative entropy in the form of a Jensen-Shannon divergence (JSD).\cite{Lin1991} The relative entropy framework has been previously established as a useful tool for evaluating the quality of coarse-grained models.\cite{chaimovich2011coarse,foley2015impact} The JSD is a variation of the well-known Kullback-Leibler divergence\cite{Kullback1951} used to calculate the relative entropy between two distributions. It offers two advantages over the Kullback-Leibler divergence in that it is symmetric and always has a finite value. Rather than directly relating two distributions, as is the case for the Kullback-Leibler divergence, the JSD computes the relative entropy by comparing each of these distributions to a third distribution which is the average of the other two distributions, as shown in the following equations

\begin{equation}
  \label{eq:jsd}
  D_{JS} = \frac{1}{2}D_{KL}\left(P_{CG}||P_{avg}\right) + \frac{1}{2}D_{KL}\left(P_{AA}||P_{avg}\right),
\end{equation}

\begin{equation*}
  \label{eq:jsd1}
  \textnormal{where } D_{KL}(A||B) = \sum_{i=1}^{N} a_i \ln\left(\frac{a_i}{b_i}\right),
\end{equation*}

\begin{equation*}
  \label{eq:jsd2}
  \textnormal{and } P_{avg} = \frac{1}{2}(P_{CG}+P_{AA}).
\end{equation*}

 Here, we use the JSD to evaluate how well the distribution of the water/octanol partition free energies for the coarse-grained molecules ($P_{CG}$) match the corresponding distribution at the atomistic resolution ($P_{AA}$). A value of 0 indicates that the two distributions are the same. The use of the average distribution ($P_{avg}$) conveniently prevents divisions by zero when comparing histograms like those shown in Fig.~\ref{fig:Hist_REs}a-d.

\subsection*{Basin Hopping and Minimization Schemes}
In this work, we use multiple methods to optimize the coarse-grained partition free energies to best match the atomistic distribution of free energies. The first such method is the basin-hopping method,\cite{wales1997global} which is a variation of Metropolis-Hastings Monte Carlo. The algorithm proceeds in the following steps. Given a set of initial coordinates and objective function, the initial coordinates are first randomly perturbed and subsequently minimized. The results of the minimization are either accepted or rejected based on a predefined Metropolis criterion. These two steps form a single iteration of the algorithm, and a large number of iterations may be required to find the desired minima. Here, we use the JSD as our objective function and a set of possible water/octanol partition free energies for each coarse-grained bead type as our initial coordinates. Each move then corresponds to shifting the values of $\dGwol$ for each coarse-grained bead type in a given force field. The optimizations were performed in order to define the desired $\dGwol$ values for the five-bead-type force field, using the basinhopping function provided by \textsc{scipy}\cite{jones2014scipy} with a Broyden-Fletcher-Goldfarb-Shanno local minimizer,\cite{nocedal2006numerical} a Metropolis temperature parameter of 0.008, and a step size of 0.024 kcal/mol. For the reference atomistic distribution, we applied the \textsc{alogps} neural network to predict $\dGwol$ for molecules in the GDB, restricting the maximum number of heavy atoms per molecule to eight. However, finding the optimal set of $\dGwol$ values for the sixteen-bead-type force field using this approach proved to be computationally unfeasible, as the dimensionality of the problem scales with $M^N$, where $N$ is the total number of bead types in the force field and $M$ is the range of $\dGwol$ values spanned by the Martini bead types divided by the step size. To parameterize the sixteen-bead-type force field, we used the \textsc{scipy} minimize function\cite{jones2014scipy} with the modified Powell method,\cite{nocedal2006numerical} starting with an initial set of eighteen bead types that were evenly distributed along the $\dGwol$ axis. The results of the minimization indicated two sets of two bead types that were within 0.1 kcal/mol of each other, and so each pair was combined into a single bead type, resulting in sixteen bead types total.

\subsection*{Clustering the GDB}
In addition to optimization of the JSD, a new set of coarse grained water/octanol partition free energies was also proposed by clustering the GDB, leading to the 9-bead-type force field. Specifically, all molecules with eight heavy atoms or less that were known to map to single bead representations using the \textsc{auto-martini} algorithm were grouped based on the number and type of hetero-atom substitutions present in the molecule (i.e., the number of times that a C was replaced with N, O, or F). The resulting atomistic molecular populations as well as the mean and standard deviation of their water/octanol partition free energies are shown in Fig.~\ref{fig:GDB_Clust}. Detailed information on each of the distributions (beyond what is provided in Fig.~\ref{fig:GDB_Clust}) is available in the SI. The desired water/octanol partition free energies are determined by clustering the points on this graph, starting from the highest populated points and accepting anything that was within plus or minus 0.5 kcal/mol of these points. For example, the first point with the highest population in Fig.~\ref{fig:GDB_Clust}a is chosen as a starting point for the first bead type. All points that fall within 0.5 kcal/mol are assigned to this bead type and the $\dGwol$ is determined by taking a population-weighted average of all of these points. The next bead type is determined by selecting the highest point on Fig.~\ref{fig:GDB_Clust}a that is not already assigned to a bead type and repeating the process.

\subsection*{Functional Group Analysis}
A statistical analysis of the functional groups found in the molecular fragments mapping to single beads is necessary in order to obtain a more detailed picture as to which chemistries are representative of specific bead types. The enumeration of functional groups was achieved through the use of the \textsc{checkmol} software developed by Haider.\cite{Haider2010} This software uses the 3D coordinates of each atom and the corresponding atom labels in a given molecule to identify common chemical functional groups. A full list of the functional groups identified can be found in the SI. Using \textsc{checkmol}, we determine the degeneracy of specific functional group pairs with respect to single bead types for the set of molecular fragments that mapped to a single bead. This amounts to counting the number of fragments containing a specific functional group pair and mapping to a single bead type. This population is then normalized with respect to the total number of fragments containing that same functional group pair across all bead types. It is useful to frame this statistical analysis in terms of conditional probabilities, as this yields specific information relevant for molecular design applications. For example, the aforementioned counting and normalization is equivalent to calculating the likelihood of assigning a bead type ($T$) given a specific functional group pair ($F$), defined as $P(T|F)$. We use the fragment population distributions for each bead type and each functional group pair to obtain probabilities $P(T)$ of a bead type and $P(F)$ of a functional group pair. We then calculate the posterior probabilities $P(F|T)$ of a given bead type back-mapping to a specific functional group pair using Bayes' theorem 

\begin{equation}
  \label{eq:bayes}
  P(F|T) = \frac{P(T|F)P(F)}{P(T)}.
\end{equation}

The results are shown as a series of heat maps for each force field in Fig.~\ref{fig:5HA_FGPairs}. 

\subsection*{Parameterization of New Bead Types}
The new force fields share most of the parameters defined by the Martini force field.\cite{Marrink2007} For the intra-molecular interactions, bonded, angle, and dihedral force constants remain the same as those prescribed by Martini. The non-bonded interactions only contain one deviation from those in Martini. We linearly interpolate across the interaction matrix defined in Martini,\cite{Marrink2007} utilizing the distance between the established Martini $\dGwol$\cite{Bereau2015} and the desired $\dGwol$ for the interpolation. The partition free energies of each bead type was then confirmed by running coarse-grained molecular dynamics simulations of single beads of each new bead type. These results are included in the SI, and show that this method yields an accurate force field without relying on an iterative scheme. Linear interpolation is chosen as it is clear that there is no underlying functional form or smooth landscape that can be derived from this parameter space (see SI for details). The new bead types are named as T$i$ types, with $i$ ranging from 1 to $N$ where $N$ is the total number of bead types in the force field. The numbering is also ordered by polarity. For example, the T1 bead type for all new force fields is the most polar type. Conversely, the T5, T9, and T16 bead types are the most apolar bead types in the five, nine, and sixteen-bead-type force fields, respectively. The full list of bead types for each force field, their force field parameters, and their corresponding $\dGwol$ values is available in the SI.

\subsection*{Coarse-Grained Simulations}
Coarse-grained molecular dynamics simulations were performed in \textsc{GROMACS}\cite{GROMACS-2008} version 4.6.6 using the standard Martini force field parameters as well as the new force field parameters derived in this work. A time step of $\delta t= 0,03~\tau$ was used for all simulations, where $\tau$ is the natural time unit for the propagation of the model defined in terms of the units of energy $\mathcal{E}$, mass $\mathcal{M}$ and length $\mathcal{L}$ as $\tau = \mathcal{L}\sqrt{\mathcal{M}/\mathcal{E}}$. The simulations were run in an $NPT$ ensemble with a Langevin thermostat and Andersen barostat\cite{Andersen1980} to keep the temperature and pressure at 300 K and 1 bar, respectively. The corresponding coupling constants were $\tau_T=\tau$ and $\tau_P=12\tau$.

Water/octanol partition free energies were obtained by simulating approximately 500 coarse-grained molecules in octanol and water. Approximately 250 octanol molecules and 350 Martini water molecules were simulated for their respective systems, with the appropriate number of antifreeze particles.\cite{Marrink2007} The free energies were computed using the Bennett acceptance ratio method\cite{bennett1976efficient} in which the coarse-grained solute was incrementally decoupled from the solvent via the coupling parameter, $\lambda$. Twenty-one simulations were run for each molecule at evenly spaced $\lambda$ values ranging from 0 to 1, with each simulation run for 200,000 time steps. Finally, the partition free energies were calculated using the relation $\Delta G_{\textup{W}\rightarrow\textup{Ol}} = \Delta G_{\textup{W}} - \Delta G_{\textup{Ol}}$.

\section*{Results}

\begin{figure}[htbp]
  \begin{center}
    \includegraphics[width=\linewidth]{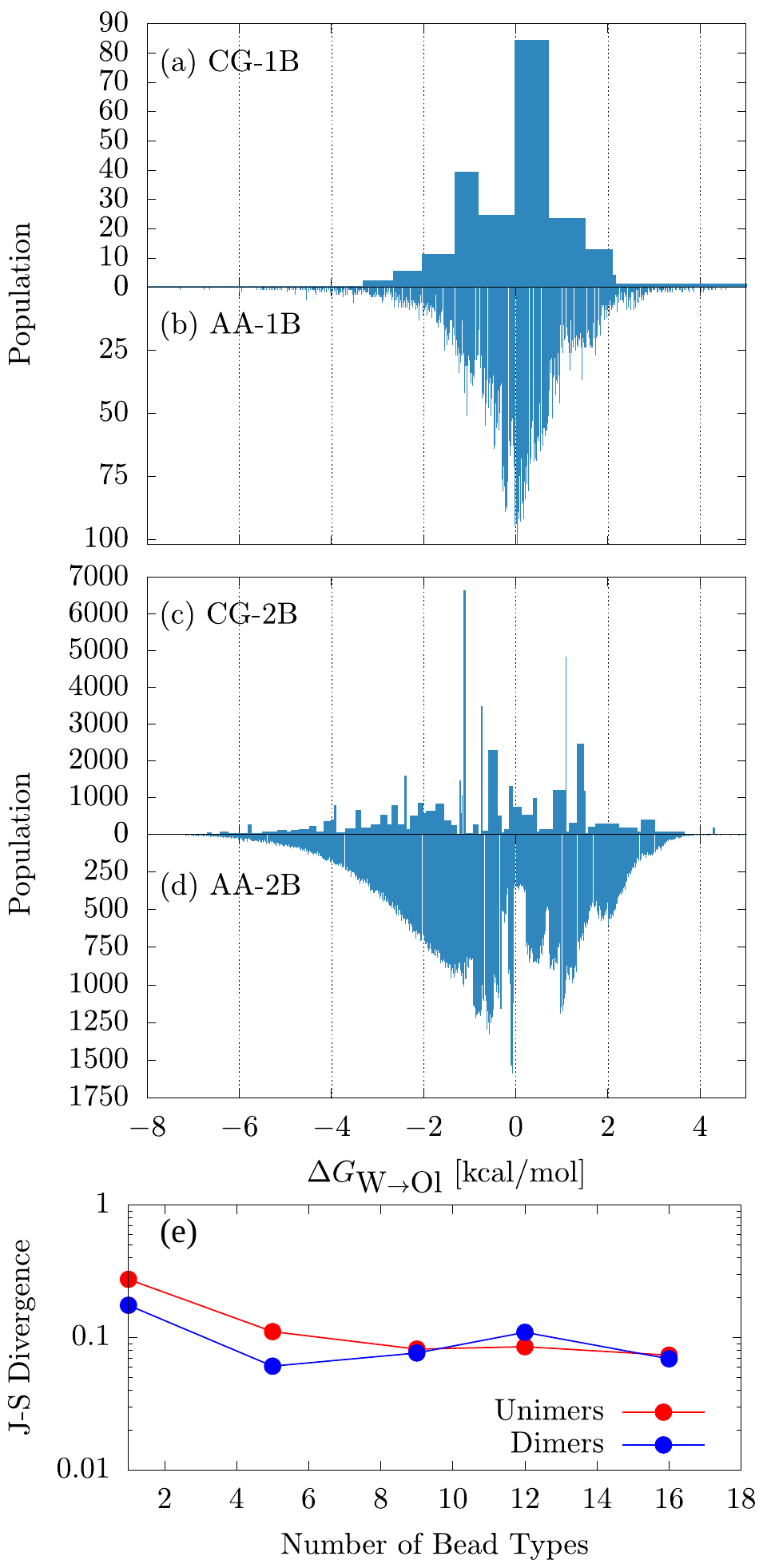}
    \caption{Histograms of 343,700 small molecules extracted from GDB that map onto one-bead or two-bead coarse-grained Martini representations. (a),(c) Coarse-grained and (b),(d) atomistic populations as a function of water/octanol partitioning free energy. The width of the bars in (a),(c) corresponds to the range of atomistic water/octanol partitioning free energies that can map to that coarse-grained representation. (e) Jensen-Shannon divergence calculated for the histograms corresponding to those in (a)-(d) for all force fields described in this work.}
    \label{fig:Hist_REs}
  \end{center}
\end{figure}

\subsection*{Quantifying information loss of CG models with varying number of bead types}

The updated \textsc{auto-martini} algorithm was used to first map and subsequently assign bead types to 3.5 million molecules of the GDB containing ten or fewer heavy atoms using the Martini force field as well as the other three force fields parameterized by interpolating the Martini interaction matrix. Fig.~\ref{fig:Hist_REs} shows a comparison of the atomistic and coarse-grained $\dGwol$ distributions for molecules mapping to Martini unimers (Fig.~\ref{fig:Hist_REs}a,b) and dimers (Fig.~\ref{fig:Hist_REs}c,d). The corresponding histograms for the other three force fields, as well as a histogram constructed using the Martini force field but with the original \textsc{auto-martini} algorithm can be found in the SI. The width of the coarse-grained bars reflects the range of $\dGwol$ values within which a molecule must fall in order to be assigned that bead type, or, in the dimer case, a combination of bead types. The height of the bars is set such that the area covered by each bar is equal to the total number of molecules that were assigned that coarse-grained representation. We then calculate the JSD between the coarse-grained and atomistic histograms for each force field to quantify the information loss as a function of the number of bead types present in each force field (Fig.~\ref{fig:Hist_REs}e). Increasing the number of bead types reduces the information loss when going from atomistic to CG resolution, though this reduction becomes insignificant after reaching nine bead types. The JSD comparing the unimer histograms (red curve in Fig.~\ref{fig:Hist_REs}e) changes negligibly when increasing the number of bead types from nine to sixteen, with only a small increase for the Martini case (12 bead types). This is expected due to the fact that the atomistic histogram of GDB molecules mapping to a single bead is a simple, unimodal distribution with a peak at $\dGwol = 0$. Since all of the force fields have at least one amphiphilic bead type with a $\dGwol$ close to 0, they all capture this defining feature of the histogram, and, comparatively, further information gains are negligible. However, the JSDs calculated from the dimer histograms (blue curve in Fig.~\ref{fig:Hist_REs}e) show a variety of interesting features. Both the nine and sixteen-bead-type force fields maintain roughly the same JSD, suggesting that the combinatorial explosion that results from doubling the molecular weight is captured by these force fields. The slight increase seen in the unimer JSD for Martini is noticeably amplified for the dimer case, indicating that careful placement of bead types on the $\dGwol$ axis is necessary to maximize chemical transferability. Surprisingly, the greatest deviation in the JSD going from the unimer to dimer histogram comes from the five-bead-type force field, dropping well below the values for the higher bead type force fields. The reason for this can be seen in Fig. S3b, which shows that the distribution of atomistic compounds mapping to dimers in the five-bead-type force field is significantly different from its analogs for the other force fields.  This indicates that a significant number of molecules that would map to dimers when using one of the other force fields are mapped to trimers or tetramers using the 5-bead-type force field.

\begin{figure*}[htbp]
  \begin{center}
    \includegraphics[width=\textwidth]{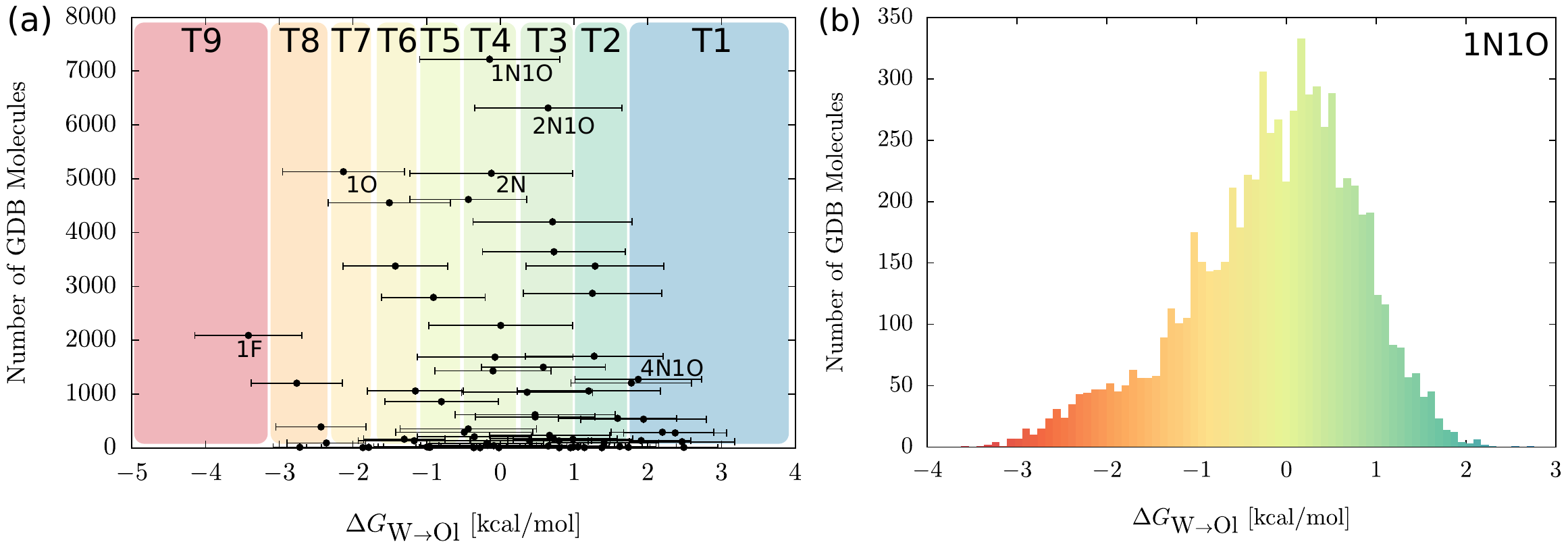}
    \caption{(a): Population versus average values of the distributions of water/octanol partition free energies corresponding to GDB molecules with up to 8 heavy atoms and a specific number and type of hetero-atom substitutions. The error bars refer to the standard deviations of each distribution. The colored backgrounds denote how these average values are clustered to obtain new bead types that more efficiently divide CCS. (b): Example distribution corresponding to top-most point labeled in the graph on the left. The label refers to the number and type of hetero-atom substitution, in this case 1 nitrogen and 1 oxygen substitution.}
    \label{fig:GDB_Clust}
  \end{center}
\end{figure*}

\subsection*{Relating chemistry to bead types}

As an alternative to purely numerical methods for determining the optimal $\dGwol$ values for the bead types of a CG force field that best partitions CCS, we cluster the GDB itself and use the weighted average of $\dGwol$ for each cluster. Fig.~\ref{fig:GDB_Clust}a shows the two descriptors upon which we project and subsequently cluster the GDB. Each point in Fig.~\ref{fig:GDB_Clust}a represents the set of molecules in the GDB that have a specific number and type of heavy atom substitutions (i.e., N, O, or F). The points are placed on the $\dGwol$ axis according to the average of their $\dGwol$ distribution. The error bars represent the standard deviation of the $\dGwol$ in each distribution. One of the corresponding distributions is shown in Fig.~\ref{fig:GDB_Clust}b. Interestingly, all of the distributions with populations of over 1000 molecules are unimodal. The points are clustered hierarchically with respect to population and average as shown in Fig.~\ref{fig:GDB_Clust}a. The highest-populated points are all chosen as cluster centers as long as they are separated by at least 0.5 kcal/mol, which is an arbitrarily chosen length-scale for the clustering to ensure a reasonable number of bead types in the final force field. After the points are clustered, the desired $\dGwol$ of each bead type is determined by taking the population-weighted average of all the points in a cluster. This intuitively provides a basic understanding of the chemistry that maps to a specific bead type. For example, a T4 bead is more likely to back-map to a molecule with one N and one O substitution compared to two N substitutions because of the difference in the GDB populations of each molecule type.  

\begin{figure*}[htbp]
  \begin{center}
    \includegraphics[height=0.9\textheight]{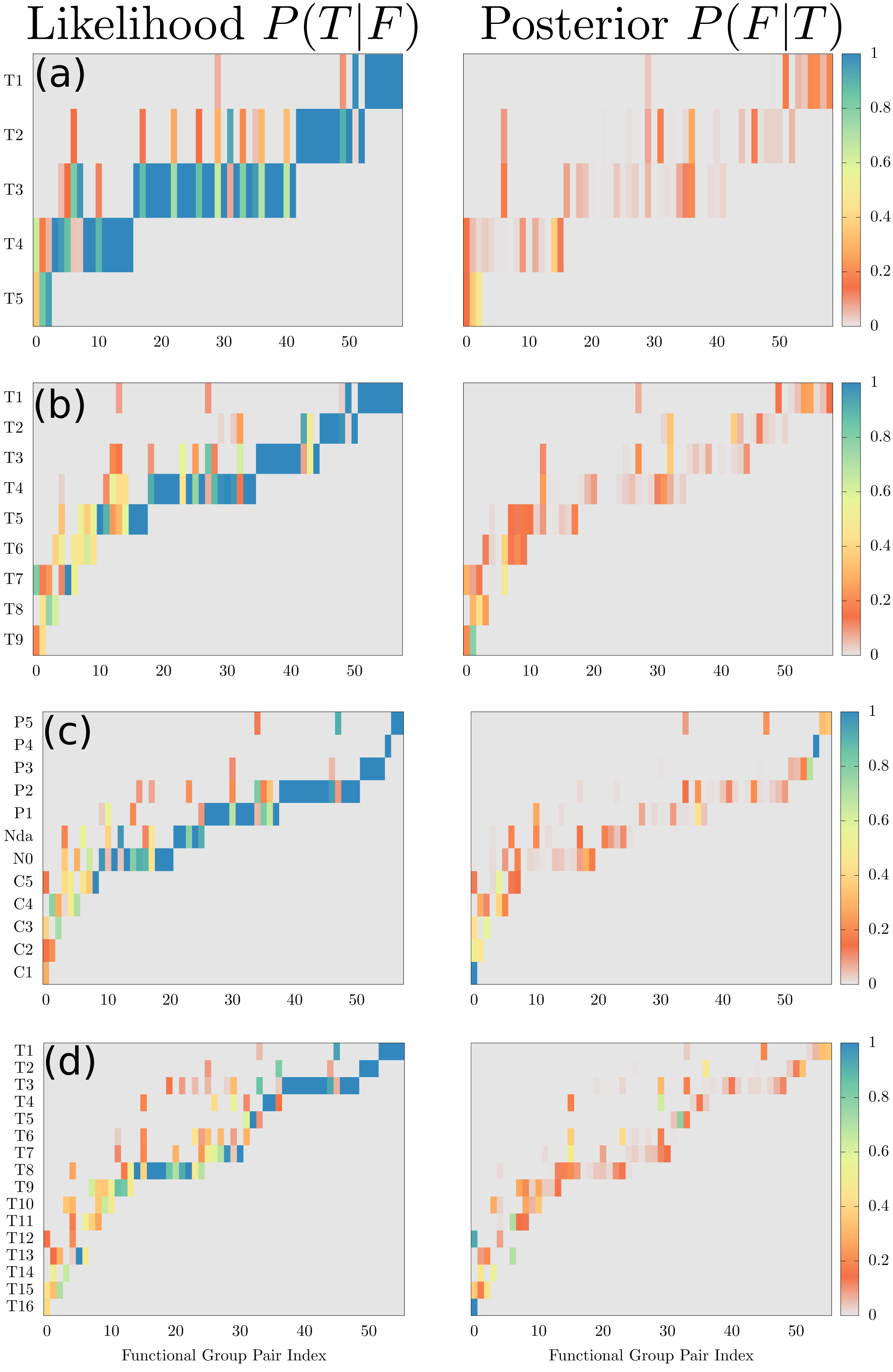}
    \caption{Heat maps portraying the degeneracy of specific pairs of functional groups for a given bead type for force fields containing five (a), nine (b), twelve (c), or sixteen (d) bead types. The horizontal axes denote specific functional group pairs that exist in a chemical fragment with five heavy atoms only. The color corresponds to either the column-normalized or row-normalized probabilities. The column-normalized probabilities (left side) are equivalent to the Bayesian likelihood of a given functional group mapping to a specific bead type. The row-normalized heat maps (right side) show the Bayesian posterior probabilities of obtaining a specific functional group given a bead type.}
    \label{fig:5HA_FGPairs}
  \end{center}
\end{figure*}

It is important to characterize the degree to which unique chemistries are captured by the bead types of each force field. Using the GDB as a proxy for CCS enables a quantitative understanding of the chemical transferability of each bead type through the calculation of conditional probabilities.  Fig.~\ref{fig:5HA_FGPairs} shows a series of heat maps corresponding to each of the four force fields investigated in this work. These heat maps are constructed by counting all fragments containing only five heavy atoms and assigned to a specific bead type, such that two functional groups are detected by the \textsc{checkmol} software package. The fragment population distributions are then used to calculate the Bayesian likelihood $P(T|F)$ and posterior $P(F|T)$ for each bead type/functional pair combination in every force field. The numbers on the horizontal axis for each heat map denote specific pairs of functional groups found in the chemical fragments that are assigned to a bead type, while the color corresponds to either the likelihood or posterior probabilities. We see the localization of functional group pairs to specific bead types mainly because of the constraint of only including fragments with five heavy atoms. This constraint limits the combinatorics of hetero-atom and bond substitutions that result in functional group pairs. Despite the addition of these constraints, a large number of functional group pairs are still split across multiple bead types. The corresponding heat maps constructed using four-heavy-atom-fragments only are included in the SI and show far less degeneracy of functional group pairs across bead types compared to these heat maps, although the general trends observed are the same. Table~\ref{table:compare} provides additional quantification of the trends seen in Fig.~\ref{fig:5HA_FGPairs}, displaying the average number of functional group pairs per bead type, as well as the number of likelihood and posterior values above cutoff values of 0.99 and 0.2, respectively. As the number of bead types increases, both the average number of functional group pairs per bead type and the number of likelihood values greater than 0.99 decrease, indicating that fewer bead types in a force field increases the coverage of CCS for each bead type. The opposite trend is observed for the number of posteriors greater than 0.2, indicating that more bead types result in higher chemical specificity for each bead type. 

\begin{table}[H]
\begin{center}
\begin{tabular}{| c | c | c | c |}
\hline
\shortstack{\# of \\Bead Types} & \shortstack{Avg. \# of \\ Func. Group Pairs}  & \shortstack{\# of\\ Likelihoods > 0.99} & \shortstack{\# of \\Posteriors > 0.20}\\
\hline
5 & 16.4 & 40 & 8\\
9 & 10.1 & 33 & 17\\
12 (Martini) & 7.4 & 35 & 20\\
16 & 6.4 & 27 & 26\\
\hline
\end{tabular}
\caption{For each force field, the number of bead types, the average number of functional group pairs per bead type, the total number of likelihood values over 0.99, and the total number of posterior values over 0.2.
}
\label{table:compare}
\end{center}
\end{table}

\begin{figure*}[htbp]
  \begin{center}
    \includegraphics[width=\linewidth]{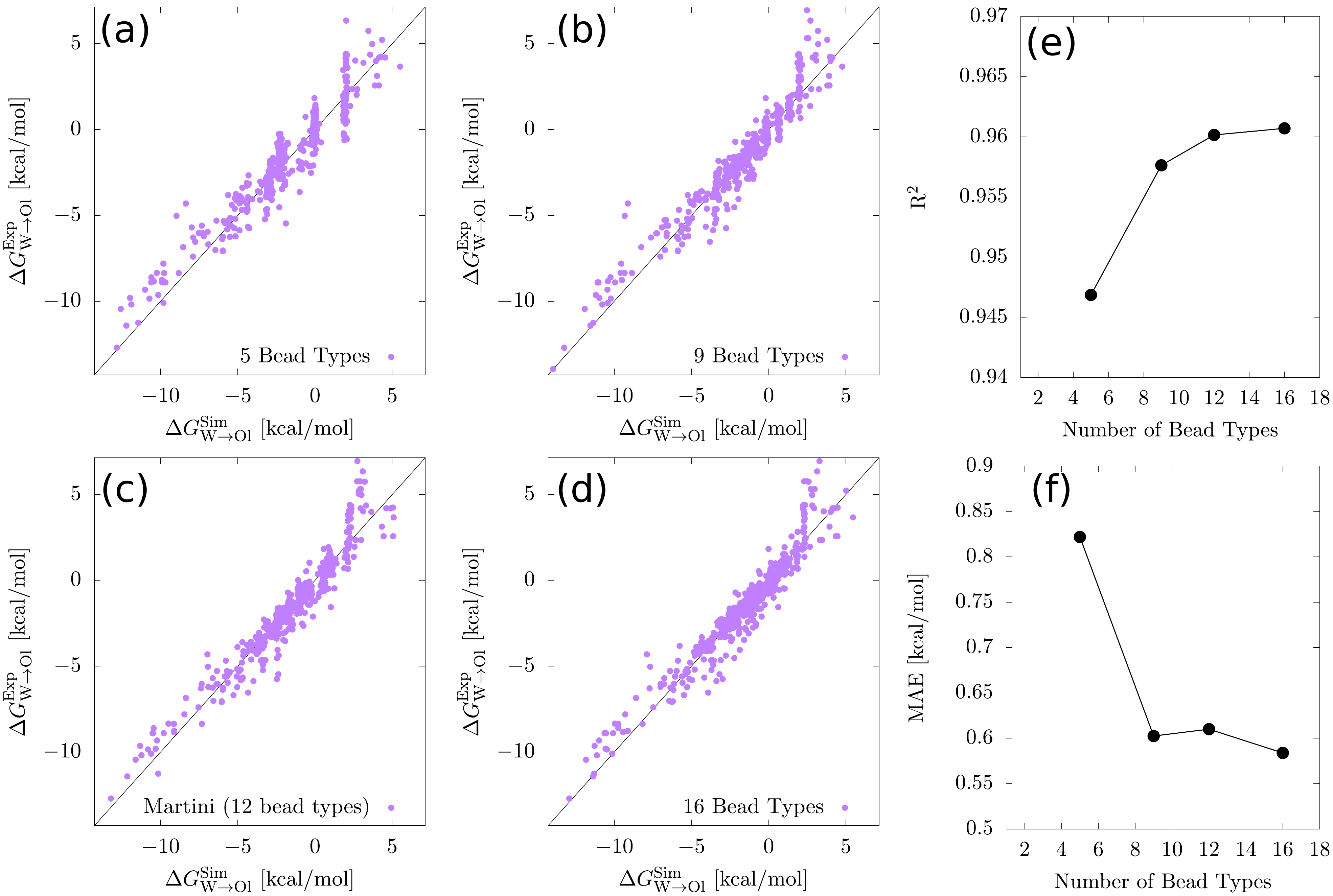}
    \caption{Correlation curves comparing the $\dGwol$ calculated from coarse-grained MD simulations of approximately 500 molecules to their measured values from experiment. The results are shown for each of the force fields described in this work (a)-(d) as well as their respective Pearson correlation coefficients and mean absolute error (MAE) values (e)-(f).}
    \label{fig:Exp_Corr}
  \end{center}
\end{figure*}

\subsection*{CG force field validation}
While we have demonstrated that the careful placement of bead types on the $\dGwol$ axis leads to more chemical transferability, the force fields themselves must be validated. Because $\dGwol$ was used as the target property for the interpolation of the Martini interaction matrix, we must ensure that this property is indeed captured by the resulting models and determine to what extent the accuracy of these models changes as the number of bead types increases. Fig.~\ref{fig:Exp_Corr} shows correlation plots comparing $\dGwol$ values computed from coarse-grained MD simulations with experimental values for approximately 500 ring-less molecules obtained from the National Cancer Institute database.\cite{voigt2001comparison} The comparison is made for all four of the models examined in this work. The number of compounds varies for each model, as the \textsc{auto-martini} algorithm was able to successfully find mappings for more molecules in the database when using a model with a higher number of bead types, ranging from 479 compounds mapped when using the five-bead-type model to 505 when using the sixteen-bead-type model. The full set of compounds as well as their corresponding coarse-grained representations is provided in the SI. The vertical series of points prominently seen in Fig.~\ref{fig:Exp_Corr}a are a consequence of the increased degeneracy of CCS for the 5-bead-type model: they represent many compounds mapping to the same coarse-grained representation. As expected, the correlation becomes less discretized as the number of bead types increases. Examining Figs.~\ref{fig:Exp_Corr}e and ~\ref{fig:Exp_Corr}f, we see corresponding gains and losses in the Pearson correlation coefficients and MAEs, respectively. Surprisingly, the gains in accuracy are very slight as a function of number of bead types---with the correlation coefficient only increasing by 0.01 and the MAE decreasing by  ~0.2 kcal/mol---despite tripling the number of bead types. Even with the five-bead-type model, we achieve an MAE of 0.8 kcal/mol, within the standard for chemical accuracy.

\section*{Discussion}

Given the immense size of CCS, the creation of reduced models that efficiently subdivide the space is necessary for screening applications. Here we demonstrate the use of the water/octanol partition free energy as the parameter used to generate top-down chemically-transferable coarse-grained models of varying numbers of bead types. This choice of descriptor is inspired by the Martini force field, which prescribes the use of $\dGwol$ when determining the bead type to be used to represent a molecular fragment. Here, we use the GDB as a proxy for CCS\cite{Menichetti2017,Menichetti2019} and apply the \textsc{auto-martini} algorithm to compare the populations of the GDB molecules and their corresponding CG representations for four different force fields with varying numbers of bead types. This effectively amounts to a discretization of CCS projected onto $\dGwol$  at multiple resolutions. Fig.~\ref{fig:Hist_REs}e quantifies the level of information loss using the JSD as the resolution is varied, allowing us to determine how effectively each of these force fields, including Martini, represents CCS. The JSD decreases as the number of bead types increase. However, the information retention becomes negligibly greater, essentially plateauing after 9 bead types. Remarkably, despite the fact that the Martini force field was parameterized using a small number of chemical compounds (relative to the large distribution of compounds used to parameterize the other models in this work), it shows only a minuscule increase in the JSD. This is mainly due to the inefficient placement of the P3, P4, and P5 beads within close proximity to each other on the $\dGwol$ axis. Unfortunately, this increase in the JSD is amplified when comparing the $\dGwol$ distributions for dimer molecules, whereas for the 9 and 16 bead type models, the JSD seems to converge. The combinatorial explosion that results from doubling the size of molecules (i.e., going from unimer to dimer) is reflected in these histograms as a broadening of the total distribution, since more hydrophobic and hydrophilic values of $\dGwol$ are possible as molecule size increases. Fig.~\ref{fig:Hist_REs}e shows that the 9 and 16 bead type force fields match this combinatorial explosion.  

On the other hand, Figs.~\ref{fig:Exp_Corr}e and f clearly demonstrate that a high level of accuracy is already achieved with respect to $\dGwol$ using the five-bead-type force field. What, then, is the benefit to using a model with more than five bead types? As we show from Figs.~\ref{fig:GDB_Clust} and ~\ref{fig:5HA_FGPairs}, the main advantage is in back-mapping the coarse-grained representations to their likely atomistic counterparts.\cite{Menichetti2018} Specifically, the 9 bead force field is parameterized not by simply optimizing the JSD, but rather by clustering the GDB molecules into sub-distributions based on the type and number of heavy-atom substitutions on the carbon scaffold of each molecule as shown in Fig.~\ref{fig:GDB_Clust}. As expected, this clustering strategy also results in a minimal value of the JSD, while providing an added convenience. The distributions that were clustered to make this force field provide a method for  predicting the chemistries that are most representative of a bead type. Since the standard deviations of these distributions are so large, such that some span across three different bead types, this provides only a rough idea of the probable chemistry accessible to a bead type. Moreover, knowledge of the presence of one or two heavy-atom substitutions on a carbon scaffold of up to 8 heavy atoms is insufficient for back-mapping given the number of ways in which they can be arranged on that scaffold resulting in wildly different chemical properties. Fig.~\ref{fig:5HA_FGPairs} shows how different functional group pairs will map clearly to specific bead types when the scaffold size is reduced to five heavy atoms. This extra constraint enables a clearer understanding of the range of unique chemistries that are accessible to a specific bead type. Decreasing the size of the scaffold from five to four heavy atoms yields correspondingly narrower distributions of $\dGwol$, meaning that the same functional group pair can be found in fewer bead types. By no longer requiring functional group pairs and increasing the scaffold size to eight heavy atoms, we begin obtaining distributions similar to those seen in Fig.~\ref{fig:GDB_Clust}. 

Table~\ref{table:compare} also demonstrates that the number of unique functional group pairs that map to a given bead type decreases as the number of bead types increases, to the point where, for Martini as well as the 16 bead type force field, there exist bead types that essentially back-map to a single functional group pair. Here, we see a clear parallel with structural coarse-graining methods: just as decreasing the size of the beads leads to a finer mapping of the configurational space, increasing the number of bead types leads to a finer mapping of CCS. The efficiency of a CG model can be optimized by tuning the mapping function and bead size of a CG model such that the accuracy of the model is balanced with respect to the computational cost of simulating a greater number of particles. By fixing the geometric mapping method and bead size, and only varying the number of bead types possible, we instead balance between the accuracy of representing specific chemical features and the cost of parameterization and validation of the inter-particle potentials. We circumvent this cost by interpolating the Martini interaction matrix to obtain accurate parameters for all of the force fields presented in this work. However, this cost will be significant for models requiring a more rigorous parameterization scheme relying on other molecular descriptors. Separate from this trade-off between accuracy and parameterization cost, a "back-mapping efficiency" can be defined as the average number of functional group pairs that map to a single bead type, indicating a larger region of CCS being captured by a single bead type. Unsurprisingly, Table ~\ref{table:compare} shows that the five-bead-type force field clearly has the highest back-mapping efficiency.

This statistical analysis of functional group pairs also suggests a Bayesian approach to computing the probability of a functional group pair, $F$, given a bead type, $T$, represented as  $P(F|T)$ in equation~\ref{eq:bayes}. $P(F)$, the Bayesian prior, is the probability of finding the specific functional group pair in the set of molecular fragments (made up of five heavy atoms and containing two functional groups as defined by \textsc{checkmol}) that mapped to single beads, and $P(T)$ is the probability of choosing the given bead type from that same data set. The likelihood, $P(T|F)$, shown in the left side of Fig.~\ref{fig:5HA_FGPairs} prescribes the bead type or types to which a fragment could be assigned based on its chemistry---the equivalent of the Martini ``bible'' for assigning bead types. As shown in Table~\ref{table:compare}, the number of functional group pairs with likelihoods greater than 0.99 (essentially localized to a single bead type) decreases as the number of bead types increases. The Martini force field deviates slightly from this trend, with two more functional group pairs with high likelihoods as compared to the nine-bead-type force field. This may stem from the parameterization strategy used for Martini that relied on specific molecules and their functional groups rather than aiming to efficiently span chemical space by optimizing the JSD, as proposed in this work. The posterior probabilities, which provide a quantitative description of which chemistries are more representative of each bead type, increase as the number of bead types increases. This effect more easily facilitates the back-mapping of coarse-grained representations. These two quantities, the Bayesian likelihood and posterior, are essential for further exploring CCS covered by specific bead types and enabling both direct and inverse molecular design.

Interestingly, we immediately see a resolution limit  with respect to the functional group pairs that map to specific bead types. Because there are certain length scales on the $\dGwol$ axis that correspond to the distribution of specific functional group pairs, increasing the number of bead types will naturally split these distributions, such that one functional group pair is represented in multiple bead types. Fig.~\ref{fig:5HA_FGPairs}a shows that the majority of functional group pairs are encompassed either by a single bead type or one of its neighbors on the $\dGwol$ axis. Increasing the number of bead types causes these splits to become more exacerbated, spanning multiple bead types for an increased number of functional group pairs.  This is the resolution limit of this type of top-down coarse-graining. The large bead sizes of these models leads to a high degree of variability in the chemistry, meaning that it is no longer obvious which functional group/functional group pair belongs to which bead type. The limit is most evident for the functional group pairs mapping to the T3 and T13 bead types in Fig.~\ref{fig:5HA_FGPairs}d, indicating that they are placed too close to their neighbors on the $\dGwol$ axis. These functional group pairs contain some combination of the following functional groups: alkene, alkyne, enamine, hydrazine, hydroxylamine, carboxylic acid derivatives, and fluorine substitution. The placement of these functional groups within a five-carbon scaffold will drastically shift the $\dGwol$ beyond the range of the next nearest bead type on the $\dGwol$ axis and highlights the limitations of only using this single descriptor for the projection of CCS. However, determining a suitable orthogonal descriptor and then parameterizing a chemically transferable CG force field to achieve a more direct relation with CCS is outside the scope of this work, and will be addressed subsequently.

\section*{Conclusion}
In this work, we use the JSD to quantify the information loss in chemically transferable top-down coarse-grained models with varying numbers of bead types, with the GDB as our proxy for chemical compound space (CCS). We find that Martini, while not designed to efficiently reduce CCS, only performs slightly worse than the other force-fields explicitly designed to minimize the JSD. All force fields yield roughly the same level of accuracy with respect to $\dGwol$, but vary greatly in their coverage of CCS. We used a Bayesian approach to calculate the probabilities of back-mapping given bead-types to fragments containing specific chemical substitutions. Here, we found it necessary to constrain the size of chemical fragments to five heavy atoms and the presence of two functional groups in order to clearly differentiate between the chemical moieties mapping to each bead type. The results of this Bayesian analysis indicate that increasing the number of bead types decreases the range of accessible chemistry while increasing the corresponding posterior probabilities for each chemistry. However, there is a resolution limit when using this approach, as it does not take into account the specific positions of hetero-atom and bond substitutions within a fragment, causing different bead types to appear representative of the same chemistry. Martini, as well as other chemically-transferable coarse-grained models, can be used to quickly build structure-property relationships that span broad regions of CCS. Here we highlight the powerful combination of this method with Bayesian inference, providing an informed mapping of a coarse structure-property relationship to a higher resolution in chemical compound space and further enabling inverse molecular design. 

\section*{Supporting Information}
The attached supporting information provides details on ($i$) the changes made to the \textsc{auto-martini} code; ($ii$) the  histograms used to calculate JSDs for all force fields described in this work; ($iii$) statistics for each of the distributions clustered when obtaining the nine-bead-type force field; ($iv$) the parameterization method for the new force fields; and ($v$) the specific functional group pairs used in the Bayesian analysis. In addition, we provide databases containing the set of GDB compounds mapping to each unimer and dimer, the force field parameters, and the trajectories simulated for each force field in a repository.~\cite{zenodo}

\begin{acknowledgments}
The authors thank Roberto Menichetti, Joseph Rudzinski, and Kurt Kremer for critical reading of the manuscript. The authors acknowledge Dr. Igor V. Tetko for an academic research license of the \textsc{alogps} software.  This work was supported by the Emmy Noether program of the Deutsche Forschungsgemeinschaft (DFG) as well as the Max Planck Graduate Center.
\end{acknowledgments}

\nocite{*}
\bibliography{biblio}% Produces the bibliography via BibTeX.

\providecommand{\noopsort}[1]{}\providecommand{\singleletter}[1]{#1}%
\begin{thebibliography}{50}%
\makeatletter
\providecommand \@ifxundefined [1]{%
 \@ifx{#1\undefined}
}%
\providecommand \@ifnum [1]{%
 \ifnum #1\expandafter \@firstoftwo
 \else \expandafter \@secondoftwo
 \fi
}%
\providecommand \@ifx [1]{%
 \ifx #1\expandafter \@firstoftwo
 \else \expandafter \@secondoftwo
 \fi
}%
\providecommand \natexlab [1]{#1}%
\providecommand \enquote  [1]{``#1''}%
\providecommand \bibnamefont  [1]{#1}%
\providecommand \bibfnamefont [1]{#1}%
\providecommand \citenamefont [1]{#1}%
\providecommand \href@noop [0]{\@secondoftwo}%
\providecommand \href [0]{\begingroup \@sanitize@url \@href}%
\providecommand \@href[1]{\@@startlink{#1}\@@href}%
\providecommand \@@href[1]{\endgroup#1\@@endlink}%
\providecommand \@sanitize@url [0]{\catcode `\\12\catcode `\$12\catcode
  `\&12\catcode `\#12\catcode `\^12\catcode `\_12\catcode `\%12\relax}%
\providecommand \@@startlink[1]{}%
\providecommand \@@endlink[0]{}%
\providecommand \url  [0]{\begingroup\@sanitize@url \@url }%
\providecommand \@url [1]{\endgroup\@href {#1}{\urlprefix }}%
\providecommand \urlprefix  [0]{URL }%
\providecommand \Eprint [0]{\href }%
\providecommand \doibase [0]{http://dx.doi.org/}%
\providecommand \selectlanguage [0]{\@gobble}%
\providecommand \bibinfo  [0]{\@secondoftwo}%
\providecommand \bibfield  [0]{\@secondoftwo}%
\providecommand \translation [1]{[#1]}%
\providecommand \BibitemOpen [0]{}%
\providecommand \bibitemStop [0]{}%
\providecommand \bibitemNoStop [0]{.\EOS\space}%
\providecommand \EOS [0]{\spacefactor3000\relax}%
\providecommand \BibitemShut  [1]{\csname bibitem#1\endcsname}%
\let\auto@bib@innerbib\@empty
%</preamble>
\bibitem [{\citenamefont {Roy}, \citenamefont {Kar},\ and\ \citenamefont
  {Das}(2015)}]{Roy2015}%
  \BibitemOpen
  \bibfield  {author} {\bibinfo {author} {\bibfnamefont {K.}~\bibnamefont
  {Roy}}, \bibinfo {author} {\bibfnamefont {S.}~\bibnamefont {Kar}}, \ and\
  \bibinfo {author} {\bibfnamefont {R.~N.}\ \bibnamefont {Das}},\ }\bibfield
  {title} {\enquote {\bibinfo {title} {{Statistical Methods in QSAR/QSPR}},}\ \
  }(\bibinfo  {publisher} {Springer, Cham},\ \bibinfo {year} {2015})\ pp.\
  \bibinfo {pages} {37--59}\BibitemShut {NoStop}%
\bibitem [{\citenamefont {Xi}\ \emph {et~al.}(2018)\citenamefont {Xi},
  \citenamefont {Pan}, \citenamefont {Li}, \citenamefont {Xu}, \citenamefont
  {Ni}, \citenamefont {Sun}, \citenamefont {Yang}, \citenamefont {Luo},
  \citenamefont {Xi}, \citenamefont {Zhu}, \citenamefont {Li}, \citenamefont
  {Jiang}, \citenamefont {Dronskowski}, \citenamefont {Shi}, \citenamefont
  {Snyder},\ and\ \citenamefont {Zhang}}]{Xi2018}%
  \BibitemOpen
  \bibfield  {author} {\bibinfo {author} {\bibfnamefont {L.}~\bibnamefont
  {Xi}}, \bibinfo {author} {\bibfnamefont {S.}~\bibnamefont {Pan}}, \bibinfo
  {author} {\bibfnamefont {X.}~\bibnamefont {Li}}, \bibinfo {author}
  {\bibfnamefont {Y.}~\bibnamefont {Xu}}, \bibinfo {author} {\bibfnamefont
  {J.}~\bibnamefont {Ni}}, \bibinfo {author} {\bibfnamefont {X.}~\bibnamefont
  {Sun}}, \bibinfo {author} {\bibfnamefont {J.}~\bibnamefont {Yang}}, \bibinfo
  {author} {\bibfnamefont {J.}~\bibnamefont {Luo}}, \bibinfo {author}
  {\bibfnamefont {J.}~\bibnamefont {Xi}}, \bibinfo {author} {\bibfnamefont
  {W.}~\bibnamefont {Zhu}}, \bibinfo {author} {\bibfnamefont {X.}~\bibnamefont
  {Li}}, \bibinfo {author} {\bibfnamefont {D.}~\bibnamefont {Jiang}}, \bibinfo
  {author} {\bibfnamefont {R.}~\bibnamefont {Dronskowski}}, \bibinfo {author}
  {\bibfnamefont {X.}~\bibnamefont {Shi}}, \bibinfo {author} {\bibfnamefont
  {G.~J.}\ \bibnamefont {Snyder}}, \ and\ \bibinfo {author} {\bibfnamefont
  {W.}~\bibnamefont {Zhang}},\ }\bibfield  {title} {\enquote {\bibinfo {title}
  {{Discovery of High Performance Thermoelectric Chalcogenides through Reliable
  High Throughput Material Screening}},}\ }\href {\doibase
  10.1021/jacs.8b04704} {\bibfield  {journal} {\bibinfo  {journal} {Journal of
  the American Chemical Society}\ }\textbf {\bibinfo {volume} {140}},\ \bibinfo
  {pages} {10785--10793} (\bibinfo {year} {2018})}\BibitemShut {NoStop}%
\bibitem [{\citenamefont {Patel}\ \emph {et~al.}(2016)\citenamefont {Patel},
  \citenamefont {Tibbitt}, \citenamefont {Celiz}, \citenamefont {Davies},
  \citenamefont {Langer}, \citenamefont {Denning}, \citenamefont {Alexander},\
  and\ \citenamefont {Anderson}}]{Patel2016}%
  \BibitemOpen
  \bibfield  {author} {\bibinfo {author} {\bibfnamefont {A.~K.}\ \bibnamefont
  {Patel}}, \bibinfo {author} {\bibfnamefont {M.~W.}\ \bibnamefont {Tibbitt}},
  \bibinfo {author} {\bibfnamefont {A.~D.}\ \bibnamefont {Celiz}}, \bibinfo
  {author} {\bibfnamefont {M.~C.}\ \bibnamefont {Davies}}, \bibinfo {author}
  {\bibfnamefont {R.}~\bibnamefont {Langer}}, \bibinfo {author} {\bibfnamefont
  {C.}~\bibnamefont {Denning}}, \bibinfo {author} {\bibfnamefont {M.~R.}\
  \bibnamefont {Alexander}}, \ and\ \bibinfo {author} {\bibfnamefont {D.~G.}\
  \bibnamefont {Anderson}},\ }\bibfield  {title} {\enquote {\bibinfo {title}
  {{High throughput screening for discovery of materials that control stem cell
  fate}},}\ }\href {\doibase 10.1016/J.COSSMS.2016.02.002} {\bibfield
  {journal} {\bibinfo  {journal} {Current Opinion in Solid State and Materials
  Science}\ }\textbf {\bibinfo {volume} {20}},\ \bibinfo {pages} {202--211}
  (\bibinfo {year} {2016})}\BibitemShut {NoStop}%
\bibitem [{\citenamefont {Mounet}\ \emph {et~al.}(2018)\citenamefont {Mounet},
  \citenamefont {Gibertini}, \citenamefont {Schwaller}, \citenamefont {Campi},
  \citenamefont {Merkys}, \citenamefont {Marrazzo}, \citenamefont {Sohier},
  \citenamefont {Castelli}, \citenamefont {Cepellotti}, \citenamefont {Pizzi}
  \emph {et~al.}}]{mounet2018two}%
  \BibitemOpen
  \bibfield  {author} {\bibinfo {author} {\bibfnamefont {N.}~\bibnamefont
  {Mounet}}, \bibinfo {author} {\bibfnamefont {M.}~\bibnamefont {Gibertini}},
  \bibinfo {author} {\bibfnamefont {P.}~\bibnamefont {Schwaller}}, \bibinfo
  {author} {\bibfnamefont {D.}~\bibnamefont {Campi}}, \bibinfo {author}
  {\bibfnamefont {A.}~\bibnamefont {Merkys}}, \bibinfo {author} {\bibfnamefont
  {A.}~\bibnamefont {Marrazzo}}, \bibinfo {author} {\bibfnamefont
  {T.}~\bibnamefont {Sohier}}, \bibinfo {author} {\bibfnamefont {I.~E.}\
  \bibnamefont {Castelli}}, \bibinfo {author} {\bibfnamefont {A.}~\bibnamefont
  {Cepellotti}}, \bibinfo {author} {\bibfnamefont {G.}~\bibnamefont {Pizzi}},
  \emph {et~al.},\ }\bibfield  {title} {\enquote {\bibinfo {title}
  {Two-dimensional materials from high-throughput computational exfoliation of
  experimentally known compounds},}\ }\href@noop {} {\bibfield  {journal}
  {\bibinfo  {journal} {Nature nanotechnology}\ }\textbf {\bibinfo {volume}
  {13}},\ \bibinfo {pages} {246} (\bibinfo {year} {2018})}\BibitemShut
  {NoStop}%
\bibitem [{\citenamefont {Greenaway}\ \emph {et~al.}(2018)\citenamefont
  {Greenaway}, \citenamefont {Santolini}, \citenamefont {Bennison},
  \citenamefont {Alston}, \citenamefont {Pugh}, \citenamefont {Little},
  \citenamefont {Miklitz}, \citenamefont {Eden-Rump}, \citenamefont {Clowes},
  \citenamefont {Shakil} \emph {et~al.}}]{greenaway2018high}%
  \BibitemOpen
  \bibfield  {author} {\bibinfo {author} {\bibfnamefont {R.}~\bibnamefont
  {Greenaway}}, \bibinfo {author} {\bibfnamefont {V.}~\bibnamefont
  {Santolini}}, \bibinfo {author} {\bibfnamefont {M.}~\bibnamefont {Bennison}},
  \bibinfo {author} {\bibfnamefont {B.}~\bibnamefont {Alston}}, \bibinfo
  {author} {\bibfnamefont {C.}~\bibnamefont {Pugh}}, \bibinfo {author}
  {\bibfnamefont {M.}~\bibnamefont {Little}}, \bibinfo {author} {\bibfnamefont
  {M.}~\bibnamefont {Miklitz}}, \bibinfo {author} {\bibfnamefont
  {E.}~\bibnamefont {Eden-Rump}}, \bibinfo {author} {\bibfnamefont
  {R.}~\bibnamefont {Clowes}}, \bibinfo {author} {\bibfnamefont
  {A.}~\bibnamefont {Shakil}},  \emph {et~al.},\ }\bibfield  {title} {\enquote
  {\bibinfo {title} {High-throughput discovery of organic cages and catenanes
  using computational screening fused with robotic synthesis},}\ }\href@noop {}
  {\bibfield  {journal} {\bibinfo  {journal} {Nature communications}\ }\textbf
  {\bibinfo {volume} {9}},\ \bibinfo {pages} {2849} (\bibinfo {year}
  {2018})}\BibitemShut {NoStop}%
\bibitem [{\citenamefont {Burns}\ \emph {et~al.}(2018)\citenamefont {Burns},
  \citenamefont {Howes}, \citenamefont {Sun}, \citenamefont {Little},
  \citenamefont {Camper}, \citenamefont {Abbott}, \citenamefont {Phan},
  \citenamefont {Lee}, \citenamefont {Waterson}, \citenamefont {Rossanese}
  \emph {et~al.}}]{burns2018high}%
  \BibitemOpen
  \bibfield  {author} {\bibinfo {author} {\bibfnamefont {M.~C.}\ \bibnamefont
  {Burns}}, \bibinfo {author} {\bibfnamefont {J.~E.}\ \bibnamefont {Howes}},
  \bibinfo {author} {\bibfnamefont {Q.}~\bibnamefont {Sun}}, \bibinfo {author}
  {\bibfnamefont {A.~J.}\ \bibnamefont {Little}}, \bibinfo {author}
  {\bibfnamefont {D.~V.}\ \bibnamefont {Camper}}, \bibinfo {author}
  {\bibfnamefont {J.~R.}\ \bibnamefont {Abbott}}, \bibinfo {author}
  {\bibfnamefont {J.}~\bibnamefont {Phan}}, \bibinfo {author} {\bibfnamefont
  {T.}~\bibnamefont {Lee}}, \bibinfo {author} {\bibfnamefont {A.~G.}\
  \bibnamefont {Waterson}}, \bibinfo {author} {\bibfnamefont {O.~W.}\
  \bibnamefont {Rossanese}},  \emph {et~al.},\ }\bibfield  {title} {\enquote
  {\bibinfo {title} {High-throughput screening identifies small molecules that
  bind to the ras: Sos: Ras complex and perturb ras signaling},}\ }\href@noop
  {} {\bibfield  {journal} {\bibinfo  {journal} {Analytical biochemistry}\
  }\textbf {\bibinfo {volume} {548}},\ \bibinfo {pages} {44--52} (\bibinfo
  {year} {2018})}\BibitemShut {NoStop}%
\bibitem [{\citenamefont {Dobson}(2004)}]{Dobson2004}%
  \BibitemOpen
  \bibfield  {author} {\bibinfo {author} {\bibfnamefont {C.~M.}\ \bibnamefont
  {Dobson}},\ }\bibfield  {title} {\enquote {\bibinfo {title} {{Chemical space
  and biology}},}\ }\href {\doibase 10.1038/nature03192} {\bibfield  {journal}
  {\bibinfo  {journal} {Nature}\ }\textbf {\bibinfo {volume} {432}},\ \bibinfo
  {pages} {824--828} (\bibinfo {year} {2004})}\BibitemShut {NoStop}%
\bibitem [{\citenamefont {Faber}\ \emph {et~al.}(2018)\citenamefont {Faber},
  \citenamefont {Christensen}, \citenamefont {Huang},\ and\ \citenamefont {von
  Lilienfeld}}]{Faber2018}%
  \BibitemOpen
  \bibfield  {author} {\bibinfo {author} {\bibfnamefont {F.~A.}\ \bibnamefont
  {Faber}}, \bibinfo {author} {\bibfnamefont {A.~S.}\ \bibnamefont
  {Christensen}}, \bibinfo {author} {\bibfnamefont {B.}~\bibnamefont {Huang}},
  \ and\ \bibinfo {author} {\bibfnamefont {O.~A.}\ \bibnamefont {von
  Lilienfeld}},\ }\bibfield  {title} {\enquote {\bibinfo {title} {{Alchemical
  and structural distribution based representation for universal quantum
  machine learning}},}\ }\href {\doibase 10.1063/1.5020710} {\bibfield
  {journal} {\bibinfo  {journal} {The Journal of Chemical Physics}\ }\textbf
  {\bibinfo {volume} {148}},\ \bibinfo {pages} {241717} (\bibinfo {year}
  {2018})}\BibitemShut {NoStop}%
\bibitem [{\citenamefont {Willatt}, \citenamefont {Musil},\ and\ \citenamefont
  {Ceriotti}(2018)}]{Willatt2018}%
  \BibitemOpen
  \bibfield  {author} {\bibinfo {author} {\bibfnamefont {M.~J.}\ \bibnamefont
  {Willatt}}, \bibinfo {author} {\bibfnamefont {F.}~\bibnamefont {Musil}}, \
  and\ \bibinfo {author} {\bibfnamefont {M.}~\bibnamefont {Ceriotti}},\
  }\bibfield  {title} {\enquote {\bibinfo {title} {{Feature optimization for
  atomistic machine learning yields a data-driven construction of the periodic
  table of the elements}},}\ }\href {\doibase 10.1039/C8CP05921G} {\bibfield
  {journal} {\bibinfo  {journal} {Physical Chemistry Chemical Physics}\
  }\textbf {\bibinfo {volume} {20}},\ \bibinfo {pages} {29661--29668} (\bibinfo
  {year} {2018})}\BibitemShut {NoStop}%
\bibitem [{\citenamefont {Ghiringhelli}\ \emph {et~al.}(2015)\citenamefont
  {Ghiringhelli}, \citenamefont {Vybiral}, \citenamefont {Levchenko},
  \citenamefont {Draxl},\ and\ \citenamefont
  {Scheffler}}]{ghiringhelli2015big}%
  \BibitemOpen
  \bibfield  {author} {\bibinfo {author} {\bibfnamefont {L.~M.}\ \bibnamefont
  {Ghiringhelli}}, \bibinfo {author} {\bibfnamefont {J.}~\bibnamefont
  {Vybiral}}, \bibinfo {author} {\bibfnamefont {S.~V.}\ \bibnamefont
  {Levchenko}}, \bibinfo {author} {\bibfnamefont {C.}~\bibnamefont {Draxl}}, \
  and\ \bibinfo {author} {\bibfnamefont {M.}~\bibnamefont {Scheffler}},\
  }\bibfield  {title} {\enquote {\bibinfo {title} {Big data of materials
  science: critical role of the descriptor},}\ }\href@noop {} {\bibfield
  {journal} {\bibinfo  {journal} {Physical review letters}\ }\textbf {\bibinfo
  {volume} {114}},\ \bibinfo {pages} {105503} (\bibinfo {year}
  {2015})}\BibitemShut {NoStop}%
\bibitem [{\citenamefont {Ramakrishnan}\ \emph {et~al.}(2015)\citenamefont
  {Ramakrishnan}, \citenamefont {Dral}, \citenamefont {Rupp},\ and\
  \citenamefont {von Lilienfeld}}]{ramakrishnan2015big}%
  \BibitemOpen
  \bibfield  {author} {\bibinfo {author} {\bibfnamefont {R.}~\bibnamefont
  {Ramakrishnan}}, \bibinfo {author} {\bibfnamefont {P.~O.}\ \bibnamefont
  {Dral}}, \bibinfo {author} {\bibfnamefont {M.}~\bibnamefont {Rupp}}, \ and\
  \bibinfo {author} {\bibfnamefont {O.~A.}\ \bibnamefont {von Lilienfeld}},\
  }\bibfield  {title} {\enquote {\bibinfo {title} {Big data meets quantum
  chemistry approximations: The $\delta$-machine learning approach},}\
  }\href@noop {} {\bibfield  {journal} {\bibinfo  {journal} {Journal of
  chemical theory and computation}\ }\textbf {\bibinfo {volume} {11}},\
  \bibinfo {pages} {2087--2096} (\bibinfo {year} {2015})}\BibitemShut {NoStop}%
\bibitem [{\citenamefont {Butler}\ \emph {et~al.}(2018)\citenamefont {Butler},
  \citenamefont {Davies}, \citenamefont {Cartwright}, \citenamefont {Isayev},\
  and\ \citenamefont {Walsh}}]{butler2018machine}%
  \BibitemOpen
  \bibfield  {author} {\bibinfo {author} {\bibfnamefont {K.~T.}\ \bibnamefont
  {Butler}}, \bibinfo {author} {\bibfnamefont {D.~W.}\ \bibnamefont {Davies}},
  \bibinfo {author} {\bibfnamefont {H.}~\bibnamefont {Cartwright}}, \bibinfo
  {author} {\bibfnamefont {O.}~\bibnamefont {Isayev}}, \ and\ \bibinfo {author}
  {\bibfnamefont {A.}~\bibnamefont {Walsh}},\ }\bibfield  {title} {\enquote
  {\bibinfo {title} {Machine learning for molecular and materials science},}\
  }\href@noop {} {\bibfield  {journal} {\bibinfo  {journal} {Nature}\ }\textbf
  {\bibinfo {volume} {559}},\ \bibinfo {pages} {547} (\bibinfo {year}
  {2018})}\BibitemShut {NoStop}%
\bibitem [{\citenamefont {Bereau}\ \emph {et~al.}(2018)\citenamefont {Bereau},
  \citenamefont {DiStasio~Jr}, \citenamefont {Tkatchenko},\ and\ \citenamefont
  {Von~Lilienfeld}}]{bereau2018non}%
  \BibitemOpen
  \bibfield  {author} {\bibinfo {author} {\bibfnamefont {T.}~\bibnamefont
  {Bereau}}, \bibinfo {author} {\bibfnamefont {R.~A.}\ \bibnamefont
  {DiStasio~Jr}}, \bibinfo {author} {\bibfnamefont {A.}~\bibnamefont
  {Tkatchenko}}, \ and\ \bibinfo {author} {\bibfnamefont {O.~A.}\ \bibnamefont
  {Von~Lilienfeld}},\ }\bibfield  {title} {\enquote {\bibinfo {title}
  {Non-covalent interactions across organic and biological subsets of chemical
  space: Physics-based potentials parametrized from machine learning},}\
  }\href@noop {} {\bibfield  {journal} {\bibinfo  {journal} {The Journal of
  chemical physics}\ }\textbf {\bibinfo {volume} {148}},\ \bibinfo {pages}
  {241706} (\bibinfo {year} {2018})}\BibitemShut {NoStop}%
\bibitem [{\citenamefont {Bereau}, \citenamefont {Andrienko},\ and\
  \citenamefont {Kremer}(2016)}]{Bereau2016}%
  \BibitemOpen
  \bibfield  {author} {\bibinfo {author} {\bibfnamefont {T.}~\bibnamefont
  {Bereau}}, \bibinfo {author} {\bibfnamefont {D.}~\bibnamefont {Andrienko}}, \
  and\ \bibinfo {author} {\bibfnamefont {K.}~\bibnamefont {Kremer}},\
  }\bibfield  {title} {\enquote {\bibinfo {title} {{Research Update:
  Computational materials discovery in soft matter}},}\ }\href@noop {}
  {\bibfield  {journal} {\bibinfo  {journal} {APL Materials}\ }\textbf
  {\bibinfo {volume} {4}},\ \bibinfo {pages} {053101} (\bibinfo {year}
  {2016})}\BibitemShut {NoStop}%
\bibitem [{\citenamefont {Bereau}(2018)}]{bereau2018data}%
  \BibitemOpen
  \bibfield  {author} {\bibinfo {author} {\bibfnamefont {T.}~\bibnamefont
  {Bereau}},\ }\bibfield  {title} {\enquote {\bibinfo {title} {Data-driven
  methods in multiscale modeling of soft matter},}\ }\href@noop {} {\bibfield
  {journal} {\bibinfo  {journal} {Handbook of Materials Modeling: Methods:
  Theory and Modeling}\ ,\ \bibinfo {pages} {1--12}} (\bibinfo {year}
  {2018})}\BibitemShut {NoStop}%
\bibitem [{\citenamefont {Noid}(2013)}]{Noid2013}%
  \BibitemOpen
  \bibfield  {author} {\bibinfo {author} {\bibfnamefont {W.~G.}\ \bibnamefont
  {Noid}},\ }\bibfield  {title} {\enquote {\bibinfo {title} {{Perspective:
  Coarse-grained models for biomolecular systems}},}\ }\href {\doibase
  10.1063/1.4818908} {\bibfield  {journal} {\bibinfo  {journal} {Journal of
  Chemical Physics}\ }\textbf {\bibinfo {volume} {139}} (\bibinfo {year}
  {2013}),\ 10.1063/1.4818908}\BibitemShut {NoStop}%
\bibitem [{\citenamefont {Kmiecik}\ \emph {et~al.}(2016)\citenamefont
  {Kmiecik}, \citenamefont {Gront}, \citenamefont {Kolinski}, \citenamefont
  {Wieteska}, \citenamefont {Dawid},\ and\ \citenamefont
  {Kolinski}}]{kmiecik2016coarse}%
  \BibitemOpen
  \bibfield  {author} {\bibinfo {author} {\bibfnamefont {S.}~\bibnamefont
  {Kmiecik}}, \bibinfo {author} {\bibfnamefont {D.}~\bibnamefont {Gront}},
  \bibinfo {author} {\bibfnamefont {M.}~\bibnamefont {Kolinski}}, \bibinfo
  {author} {\bibfnamefont {L.}~\bibnamefont {Wieteska}}, \bibinfo {author}
  {\bibfnamefont {A.~E.}\ \bibnamefont {Dawid}}, \ and\ \bibinfo {author}
  {\bibfnamefont {A.}~\bibnamefont {Kolinski}},\ }\bibfield  {title} {\enquote
  {\bibinfo {title} {Coarse-grained protein models and their applications},}\
  }\href@noop {} {\bibfield  {journal} {\bibinfo  {journal} {Chemical reviews}\
  }\textbf {\bibinfo {volume} {116}},\ \bibinfo {pages} {7898--7936} (\bibinfo
  {year} {2016})}\BibitemShut {NoStop}%
\bibitem [{\citenamefont {Voth}(2008)}]{voth2008coarse}%
  \BibitemOpen
  \bibfield  {author} {\bibinfo {author} {\bibfnamefont {G.~A.}\ \bibnamefont
  {Voth}},\ }\href@noop {} {\emph {\bibinfo {title} {Coarse-graining of
  condensed phase and biomolecular systems}}}\ (\bibinfo  {publisher} {CRC
  press},\ \bibinfo {year} {2008})\BibitemShut {NoStop}%
\bibitem [{\citenamefont {Peter}\ and\ \citenamefont
  {Kremer}(2009)}]{peter2009multiscale}%
  \BibitemOpen
  \bibfield  {author} {\bibinfo {author} {\bibfnamefont {C.}~\bibnamefont
  {Peter}}\ and\ \bibinfo {author} {\bibfnamefont {K.}~\bibnamefont {Kremer}},\
  }\bibfield  {title} {\enquote {\bibinfo {title} {Multiscale simulation of
  soft matter systems--from the atomistic to the coarse-grained level and
  back},}\ }\href@noop {} {\bibfield  {journal} {\bibinfo  {journal} {Soft
  Matter}\ }\textbf {\bibinfo {volume} {5}},\ \bibinfo {pages} {4357--4366}
  (\bibinfo {year} {2009})}\BibitemShut {NoStop}%
\bibitem [{\citenamefont {Schommers}(1983)}]{schommers1983pair}%
  \BibitemOpen
  \bibfield  {author} {\bibinfo {author} {\bibfnamefont {W.}~\bibnamefont
  {Schommers}},\ }\bibfield  {title} {\enquote {\bibinfo {title} {Pair
  potentials in disordered many-particle systems: A study for liquid
  gallium},}\ }\href@noop {} {\bibfield  {journal} {\bibinfo  {journal}
  {Physical Review A}\ }\textbf {\bibinfo {volume} {28}},\ \bibinfo {pages}
  {3599} (\bibinfo {year} {1983})}\BibitemShut {NoStop}%
\bibitem [{\citenamefont {Tsch{\"o}p}\ \emph {et~al.}(1998)\citenamefont
  {Tsch{\"o}p}, \citenamefont {Kremer}, \citenamefont {Batoulis}, \citenamefont
  {B{\"u}rger},\ and\ \citenamefont {Hahn}}]{tschop1998simulation}%
  \BibitemOpen
  \bibfield  {author} {\bibinfo {author} {\bibfnamefont {W.}~\bibnamefont
  {Tsch{\"o}p}}, \bibinfo {author} {\bibfnamefont {K.}~\bibnamefont {Kremer}},
  \bibinfo {author} {\bibfnamefont {J.}~\bibnamefont {Batoulis}}, \bibinfo
  {author} {\bibfnamefont {T.}~\bibnamefont {B{\"u}rger}}, \ and\ \bibinfo
  {author} {\bibfnamefont {O.}~\bibnamefont {Hahn}},\ }\bibfield  {title}
  {\enquote {\bibinfo {title} {Simulation of polymer melts. i. coarse-graining
  procedure for polycarbonates},}\ }\href@noop {} {\bibfield  {journal}
  {\bibinfo  {journal} {Acta Polymerica}\ }\textbf {\bibinfo {volume} {49}},\
  \bibinfo {pages} {61--74} (\bibinfo {year} {1998})}\BibitemShut {NoStop}%
\bibitem [{\citenamefont {Ercolessi}\ and\ \citenamefont
  {Adams}(1994)}]{Ercolessi1994}%
  \BibitemOpen
  \bibfield  {author} {\bibinfo {author} {\bibfnamefont {F.}~\bibnamefont
  {Ercolessi}}\ and\ \bibinfo {author} {\bibfnamefont {J.~B.}\ \bibnamefont
  {Adams}},\ }\bibfield  {title} {\enquote {\bibinfo {title} {{Interatomic
  Potentials from First-Principles Calculations: The Force-Matching Method}},}\
  }\href {\doibase 10.1209/0295-5075/26/8/005} {\bibfield  {journal} {\bibinfo
  {journal} {Europhysics Letters (EPL)}\ }\textbf {\bibinfo {volume} {26}},\
  \bibinfo {pages} {583--588} (\bibinfo {year} {1994})}\BibitemShut {NoStop}%
\bibitem [{\citenamefont {Izvekov}\ and\ \citenamefont
  {Voth}(2005)}]{izvekov2005multiscale}%
  \BibitemOpen
  \bibfield  {author} {\bibinfo {author} {\bibfnamefont {S.}~\bibnamefont
  {Izvekov}}\ and\ \bibinfo {author} {\bibfnamefont {G.~A.}\ \bibnamefont
  {Voth}},\ }\bibfield  {title} {\enquote {\bibinfo {title} {A multiscale
  coarse-graining method for biomolecular systems},}\ }\href@noop {} {\bibfield
   {journal} {\bibinfo  {journal} {The Journal of Physical Chemistry B}\
  }\textbf {\bibinfo {volume} {109}},\ \bibinfo {pages} {2469--2473} (\bibinfo
  {year} {2005})}\BibitemShut {NoStop}%
\bibitem [{\citenamefont {Ayton}, \citenamefont {Noid},\ and\ \citenamefont
  {Voth}(2007)}]{ayton2007multiscale}%
  \BibitemOpen
  \bibfield  {author} {\bibinfo {author} {\bibfnamefont {G.~S.}\ \bibnamefont
  {Ayton}}, \bibinfo {author} {\bibfnamefont {W.~G.}\ \bibnamefont {Noid}}, \
  and\ \bibinfo {author} {\bibfnamefont {G.~A.}\ \bibnamefont {Voth}},\
  }\bibfield  {title} {\enquote {\bibinfo {title} {Multiscale modeling of
  biomolecular systems: in serial and in parallel},}\ }\href@noop {} {\bibfield
   {journal} {\bibinfo  {journal} {Current opinion in structural biology}\
  }\textbf {\bibinfo {volume} {17}},\ \bibinfo {pages} {192--198} (\bibinfo
  {year} {2007})}\BibitemShut {NoStop}%
\bibitem [{\citenamefont {Marrink}, \citenamefont {de~Vries},\ and\
  \citenamefont {Mark}(2004)}]{M-2004}%
  \BibitemOpen
  \bibfield  {author} {\bibinfo {author} {\bibfnamefont {S.~J.}\ \bibnamefont
  {Marrink}}, \bibinfo {author} {\bibfnamefont {A.~H.}\ \bibnamefont
  {de~Vries}}, \ and\ \bibinfo {author} {\bibfnamefont {A.~E.}\ \bibnamefont
  {Mark}},\ }\bibfield  {title} {\enquote {\bibinfo {title} {{Coarse Grained
  Model for Semiquantitative Lipid Simulations}},}\ }\href@noop {} {\bibfield
  {journal} {\bibinfo  {journal} {J. Phys. Chem. B}\ }\textbf {\bibinfo
  {volume} {108}},\ \bibinfo {pages} {750--760} (\bibinfo {year}
  {2004})}\BibitemShut {NoStop}%
\bibitem [{\citenamefont {Monticelli}\ \emph {et~al.}(2008)\citenamefont
  {Monticelli}, \citenamefont {Kandasamy}, \citenamefont {Periole},
  \citenamefont {Larson}, \citenamefont {Tieleman},\ and\ \citenamefont
  {Marrink}}]{M-2008}%
  \BibitemOpen
  \bibfield  {author} {\bibinfo {author} {\bibfnamefont {L.}~\bibnamefont
  {Monticelli}}, \bibinfo {author} {\bibfnamefont {S.~K.}\ \bibnamefont
  {Kandasamy}}, \bibinfo {author} {\bibfnamefont {X.}~\bibnamefont {Periole}},
  \bibinfo {author} {\bibfnamefont {R.~G.}\ \bibnamefont {Larson}}, \bibinfo
  {author} {\bibfnamefont {D.~P.}\ \bibnamefont {Tieleman}}, \ and\ \bibinfo
  {author} {\bibfnamefont {S.-J.}\ \bibnamefont {Marrink}},\ }\bibfield
  {title} {\enquote {\bibinfo {title} {The martini coarse-grained force field:
  extension to proteins},}\ }\href@noop {} {\bibfield  {journal} {\bibinfo
  {journal} {Journal of chemical theory and computation}\ }\textbf {\bibinfo
  {volume} {4}},\ \bibinfo {pages} {819--834} (\bibinfo {year}
  {2008})}\BibitemShut {NoStop}%
\bibitem [{\citenamefont {Marrink}\ and\ \citenamefont
  {Tieleman}(2013)}]{Marrink2013}%
  \BibitemOpen
  \bibfield  {author} {\bibinfo {author} {\bibfnamefont {S.~J.}\ \bibnamefont
  {Marrink}}\ and\ \bibinfo {author} {\bibfnamefont {D.~P.}\ \bibnamefont
  {Tieleman}},\ }\bibfield  {title} {\enquote {\bibinfo {title} {{Perspective
  on the Martini model.}}}\ }\href {\doibase 10.1039/c3cs60093a} {\bibfield
  {journal} {\bibinfo  {journal} {Chemical Society reviews}\ }\textbf {\bibinfo
  {volume} {42}},\ \bibinfo {pages} {6801--6822} (\bibinfo {year}
  {2013})}\BibitemShut {NoStop}%
\bibitem [{\citenamefont {Menichetti}\ \emph {et~al.}(2017)\citenamefont
  {Menichetti}, \citenamefont {Kanekal}, \citenamefont {Kremer},\ and\
  \citenamefont {Bereau}}]{Menichetti2017}%
  \BibitemOpen
  \bibfield  {author} {\bibinfo {author} {\bibfnamefont {R.}~\bibnamefont
  {Menichetti}}, \bibinfo {author} {\bibfnamefont {K.~H.}\ \bibnamefont
  {Kanekal}}, \bibinfo {author} {\bibfnamefont {K.}~\bibnamefont {Kremer}}, \
  and\ \bibinfo {author} {\bibfnamefont {T.}~\bibnamefont {Bereau}},\
  }\bibfield  {title} {\enquote {\bibinfo {title} {{In silico screening of
  drug-membrane thermodynamics reveals linear relations between bulk
  partitioning and the potential of mean force}},}\ }\href {\doibase
  10.1063/1.4987012} {\bibfield  {journal} {\bibinfo  {journal} {The Journal of
  Chemical Physics}\ }\textbf {\bibinfo {volume} {147}},\ \bibinfo {pages}
  {125101} (\bibinfo {year} {2017})}\BibitemShut {NoStop}%
\bibitem [{\citenamefont {Menichetti}, \citenamefont {Kanekal},\ and\
  \citenamefont {Bereau}(2019)}]{Menichetti2019}%
  \BibitemOpen
  \bibfield  {author} {\bibinfo {author} {\bibfnamefont {R.}~\bibnamefont
  {Menichetti}}, \bibinfo {author} {\bibfnamefont {K.~H.}\ \bibnamefont
  {Kanekal}}, \ and\ \bibinfo {author} {\bibfnamefont {T.}~\bibnamefont
  {Bereau}},\ }\bibfield  {title} {\enquote {\bibinfo {title} {{Drug–Membrane
  Permeability across Chemical Space}},}\ }\href {\doibase
  10.1021/acscentsci.8b00718} {\bibfield  {journal} {\bibinfo  {journal} {ACS
  Central Science}\ }\textbf {\bibinfo {volume} {5}},\ \bibinfo {pages}
  {290--298} (\bibinfo {year} {2019})}\BibitemShut {NoStop}%
\bibitem [{\citenamefont {Fink}, \citenamefont {Bruggesser},\ and\
  \citenamefont {Reymond}(2005)}]{Fink2005}%
  \BibitemOpen
  \bibfield  {author} {\bibinfo {author} {\bibfnamefont {T.}~\bibnamefont
  {Fink}}, \bibinfo {author} {\bibfnamefont {H.}~\bibnamefont {Bruggesser}}, \
  and\ \bibinfo {author} {\bibfnamefont {J.~L.}\ \bibnamefont {Reymond}},\
  }\bibfield  {title} {\enquote {\bibinfo {title} {{Virtual exploration of the
  small-molecule chemical universe below 160 daltons}},}\ }\href {\doibase
  10.1002/anie.200462457} {\bibfield  {journal} {\bibinfo  {journal}
  {Angewandte Chemie - International Edition}\ }\textbf {\bibinfo {volume}
  {44}},\ \bibinfo {pages} {1504--1508} (\bibinfo {year} {2005})}\BibitemShut
  {NoStop}%
\bibitem [{\citenamefont {Fink}\ and\ \citenamefont
  {Reymond}(2007)}]{fink2007virtual}%
  \BibitemOpen
  \bibfield  {author} {\bibinfo {author} {\bibfnamefont {T.}~\bibnamefont
  {Fink}}\ and\ \bibinfo {author} {\bibfnamefont {J.-L.}\ \bibnamefont
  {Reymond}},\ }\bibfield  {title} {\enquote {\bibinfo {title} {Virtual
  exploration of the chemical universe up to 11 atoms of c, n, o, f: assembly
  of 26.4 million structures (110.9 million stereoisomers) and analysis for new
  ring systems, stereochemistry, physicochemical properties, compound classes,
  and drug discovery},}\ }\href@noop {} {\bibfield  {journal} {\bibinfo
  {journal} {Journal of chemical information and modeling}\ }\textbf {\bibinfo
  {volume} {47}},\ \bibinfo {pages} {342--353} (\bibinfo {year}
  {2007})}\BibitemShut {NoStop}%
\bibitem [{\citenamefont {Bereau}\ and\ \citenamefont
  {Kremer}(2015)}]{Bereau2015}%
  \BibitemOpen
  \bibfield  {author} {\bibinfo {author} {\bibfnamefont {T.}~\bibnamefont
  {Bereau}}\ and\ \bibinfo {author} {\bibfnamefont {K.}~\bibnamefont
  {Kremer}},\ }\bibfield  {title} {\enquote {\bibinfo {title} {Automated
  parametrization of the coarse-grained martini force field for small organic
  molecules},}\ }\href@noop {} {\bibfield  {journal} {\bibinfo  {journal}
  {Journal of chemical theory and computation}\ }\textbf {\bibinfo {volume}
  {11}},\ \bibinfo {pages} {2783--2791} (\bibinfo {year} {2015})}\BibitemShut
  {NoStop}%
\bibitem [{\citenamefont {Tetko}, \citenamefont {Tanchuk},\ and\ \citenamefont
  {Villa}(2001)}]{Tetko2001}%
  \BibitemOpen
  \bibfield  {author} {\bibinfo {author} {\bibfnamefont {I.~V.}\ \bibnamefont
  {Tetko}}, \bibinfo {author} {\bibfnamefont {V.~Y.}\ \bibnamefont {Tanchuk}},
  \ and\ \bibinfo {author} {\bibfnamefont {a.~E.}\ \bibnamefont {Villa}},\
  }\bibfield  {title} {\enquote {\bibinfo {title} {{Prediction of
  n-octanol/water partition coefficients from PHYSPROP database using
  artificial neural networks and E-state indices.}}}\ }\href {\doibase
  10.1021/ci010368v} {\bibfield  {journal} {\bibinfo  {journal} {Journal of
  chemical information and computer sciences}\ }\textbf {\bibinfo {volume}
  {41}},\ \bibinfo {pages} {1407--1421} (\bibinfo {year} {2001})}\BibitemShut
  {NoStop}%
\bibitem [{\citenamefont {Tetko}\ and\ \citenamefont
  {Tanchuk}(2002)}]{Tetko2002}%
  \BibitemOpen
  \bibfield  {author} {\bibinfo {author} {\bibfnamefont {I.~V.}\ \bibnamefont
  {Tetko}}\ and\ \bibinfo {author} {\bibfnamefont {V.~Y.}\ \bibnamefont
  {Tanchuk}},\ }\bibfield  {title} {\enquote {\bibinfo {title} {{Application of
  associative neural networks for prediction of lipophilicity in ALOGPS 2.1
  program}},}\ }\href {\doibase 10.1021/ci025515j} {\bibfield  {journal}
  {\bibinfo  {journal} {Journal of Chemical Information and Computer Sciences}\
  }\textbf {\bibinfo {volume} {42}},\ \bibinfo {pages} {1136--1145} (\bibinfo
  {year} {2002})}\BibitemShut {NoStop}%
\bibitem [{\citenamefont {Oliphant}(2006)}]{oliphant2006guide}%
  \BibitemOpen
  \bibfield  {author} {\bibinfo {author} {\bibfnamefont {T.~E.}\ \bibnamefont
  {Oliphant}},\ }\href@noop {} {\emph {\bibinfo {title} {A guide to NumPy}}},\
  Vol.~\bibinfo {volume} {1}\ (\bibinfo  {publisher} {Trelgol Publishing USA},\
  \bibinfo {year} {2006})\BibitemShut {NoStop}%
\bibitem [{\citenamefont {Lin}(1991)}]{Lin1991}%
  \BibitemOpen
  \bibfield  {author} {\bibinfo {author} {\bibfnamefont {J.}~\bibnamefont
  {Lin}},\ }\bibfield  {title} {\enquote {\bibinfo {title} {Divergence measures
  based on the shannon entropy},}\ }\href@noop {} {\bibfield  {journal}
  {\bibinfo  {journal} {IEEE Transactions on Information theory}\ }\textbf
  {\bibinfo {volume} {37}},\ \bibinfo {pages} {145--151} (\bibinfo {year}
  {1991})}\BibitemShut {NoStop}%
\bibitem [{\citenamefont {Chaimovich}\ and\ \citenamefont
  {Shell}(2011)}]{chaimovich2011coarse}%
  \BibitemOpen
  \bibfield  {author} {\bibinfo {author} {\bibfnamefont {A.}~\bibnamefont
  {Chaimovich}}\ and\ \bibinfo {author} {\bibfnamefont {M.~S.}\ \bibnamefont
  {Shell}},\ }\bibfield  {title} {\enquote {\bibinfo {title} {Coarse-graining
  errors and numerical optimization using a relative entropy framework},}\
  }\href@noop {} {\bibfield  {journal} {\bibinfo  {journal} {The Journal of
  chemical physics}\ }\textbf {\bibinfo {volume} {134}},\ \bibinfo {pages}
  {094112} (\bibinfo {year} {2011})}\BibitemShut {NoStop}%
\bibitem [{\citenamefont {Foley}, \citenamefont {Shell},\ and\ \citenamefont
  {Noid}(2015)}]{foley2015impact}%
  \BibitemOpen
  \bibfield  {author} {\bibinfo {author} {\bibfnamefont {T.~T.}\ \bibnamefont
  {Foley}}, \bibinfo {author} {\bibfnamefont {M.~S.}\ \bibnamefont {Shell}}, \
  and\ \bibinfo {author} {\bibfnamefont {W.~G.}\ \bibnamefont {Noid}},\
  }\bibfield  {title} {\enquote {\bibinfo {title} {The impact of resolution
  upon entropy and information in coarse-grained models},}\ }\href@noop {}
  {\bibfield  {journal} {\bibinfo  {journal} {The Journal of chemical physics}\
  }\textbf {\bibinfo {volume} {143}},\ \bibinfo {pages} {12B601\_1} (\bibinfo
  {year} {2015})}\BibitemShut {NoStop}%
\bibitem [{\citenamefont {Kullback}\ and\ \citenamefont
  {Leibler}(1951)}]{Kullback1951}%
  \BibitemOpen
  \bibfield  {author} {\bibinfo {author} {\bibfnamefont {S.}~\bibnamefont
  {Kullback}}\ and\ \bibinfo {author} {\bibfnamefont {R.~A.}\ \bibnamefont
  {Leibler}},\ }\bibfield  {title} {\enquote {\bibinfo {title} {{On Information
  and Sufficiency}},}\ }\href {\doibase 10.1214/aoms/1177729694} {\bibfield
  {journal} {\bibinfo  {journal} {The Annals of Mathematical Statistics}\
  }\textbf {\bibinfo {volume} {22}},\ \bibinfo {pages} {79--86} (\bibinfo
  {year} {1951})}\BibitemShut {NoStop}%
\bibitem [{\citenamefont {Wales}\ and\ \citenamefont
  {Doye}(1997)}]{wales1997global}%
  \BibitemOpen
  \bibfield  {author} {\bibinfo {author} {\bibfnamefont {D.~J.}\ \bibnamefont
  {Wales}}\ and\ \bibinfo {author} {\bibfnamefont {J.~P.}\ \bibnamefont
  {Doye}},\ }\bibfield  {title} {\enquote {\bibinfo {title} {Global
  optimization by basin-hopping and the lowest energy structures of
  lennard-jones clusters containing up to 110 atoms},}\ }\href@noop {}
  {\bibfield  {journal} {\bibinfo  {journal} {The Journal of Physical Chemistry
  A}\ }\textbf {\bibinfo {volume} {101}},\ \bibinfo {pages} {5111--5116}
  (\bibinfo {year} {1997})}\BibitemShut {NoStop}%
\bibitem [{\citenamefont {Jones}, \citenamefont {Oliphant},\ and\ \citenamefont
  {Peterson}(2014)}]{jones2014scipy}%
  \BibitemOpen
  \bibfield  {author} {\bibinfo {author} {\bibfnamefont {E.}~\bibnamefont
  {Jones}}, \bibinfo {author} {\bibfnamefont {T.}~\bibnamefont {Oliphant}}, \
  and\ \bibinfo {author} {\bibfnamefont {P.}~\bibnamefont {Peterson}},\
  }\bibfield  {title} {\enquote {\bibinfo {title} {$\{$SciPy$\}$: Open source
  scientific tools for $\{$Python$\}$},}\ }\href@noop {} {\  (\bibinfo {year}
  {2014})}\BibitemShut {NoStop}%
\bibitem [{\citenamefont {Nocedal}\ and\ \citenamefont
  {Wright}(2006)}]{nocedal2006numerical}%
  \BibitemOpen
  \bibfield  {author} {\bibinfo {author} {\bibfnamefont {J.}~\bibnamefont
  {Nocedal}}\ and\ \bibinfo {author} {\bibfnamefont {S.}~\bibnamefont
  {Wright}},\ }\href@noop {} {\emph {\bibinfo {title} {Numerical
  optimization}}}\ (\bibinfo  {publisher} {Springer Science \& Business
  Media},\ \bibinfo {year} {2006})\BibitemShut {NoStop}%
\bibitem [{\citenamefont {Haider}(2010)}]{Haider2010}%
  \BibitemOpen
  \bibfield  {author} {\bibinfo {author} {\bibfnamefont {N.}~\bibnamefont
  {Haider}},\ }\bibfield  {title} {\enquote {\bibinfo {title} {{Functionality
  Pattern Matching as an Efficient Complementary Structure/Reaction Search
  Tool: an Open-Source Approach}},}\ }\href {\doibase
  10.3390/molecules15085079} {\bibfield  {journal} {\bibinfo  {journal}
  {Molecules}\ }\textbf {\bibinfo {volume} {15}},\ \bibinfo {pages}
  {5079--5092} (\bibinfo {year} {2010})}\BibitemShut {NoStop}%
\bibitem [{\citenamefont {Marrink}\ \emph {et~al.}(2007)\citenamefont
  {Marrink}, \citenamefont {Risselada}, \citenamefont {Yefimov}, \citenamefont
  {Tieleman},\ and\ \citenamefont {Vries}}]{Marrink2007}%
  \BibitemOpen
  \bibfield  {author} {\bibinfo {author} {\bibfnamefont {S.~J.}\ \bibnamefont
  {Marrink}}, \bibinfo {author} {\bibfnamefont {H.~J.}\ \bibnamefont
  {Risselada}}, \bibinfo {author} {\bibfnamefont {S.}~\bibnamefont {Yefimov}},
  \bibinfo {author} {\bibfnamefont {D.~P.}\ \bibnamefont {Tieleman}}, \ and\
  \bibinfo {author} {\bibfnamefont {A.~H.~D.}\ \bibnamefont {Vries}},\
  }\bibfield  {title} {\enquote {\bibinfo {title} {{The MARTINI Force Field :
  Coarse Grained Model for Biomolecular Simulations The MARTINI Force Field :
  Coarse Grained Model for Biomolecular Simulations}},}\ }\href {\doibase
  10.1021/jp071097f} {\bibfield  {journal} {\bibinfo  {journal} {Journal of
  Physical Chemistry B}\ }\textbf {\bibinfo {volume} {111}},\ \bibinfo {pages}
  {7812--7824} (\bibinfo {year} {2007})}\BibitemShut {NoStop}%
\bibitem [{\citenamefont {Hess}\ \emph {et~al.}(2008)\citenamefont {Hess},
  \citenamefont {Kutzner}, \citenamefont {{Van Der Spoel}},\ and\ \citenamefont
  {Lindahl}}]{GROMACS-2008}%
  \BibitemOpen
  \bibfield  {author} {\bibinfo {author} {\bibfnamefont {B.}~\bibnamefont
  {Hess}}, \bibinfo {author} {\bibfnamefont {C.}~\bibnamefont {Kutzner}},
  \bibinfo {author} {\bibfnamefont {D.}~\bibnamefont {{Van Der Spoel}}}, \ and\
  \bibinfo {author} {\bibfnamefont {E.}~\bibnamefont {Lindahl}},\ }\bibfield
  {title} {\enquote {\bibinfo {title} {{GROMACS 4: Algorithms for Highly
  Efficient, Load-Balanced, and Scalable Mol. Simulat.}}}\ }\href@noop {}
  {\bibfield  {journal} {\bibinfo  {journal} {J. Chem. Theory Comput.}\
  }\textbf {\bibinfo {volume} {4}},\ \bibinfo {pages} {435−447} (\bibinfo
  {year} {2008})}\BibitemShut {NoStop}%
\bibitem [{\citenamefont {Andersen}(1980)}]{Andersen1980}%
  \BibitemOpen
  \bibfield  {author} {\bibinfo {author} {\bibfnamefont {H.~C.}\ \bibnamefont
  {Andersen}},\ }\bibfield  {title} {\enquote {\bibinfo {title} {{Molecular
  dynamics simulations at constant pressure and/or temperature}},}\ }\href
  {\doibase 10.1063/1.439486} {\bibfield  {journal} {\bibinfo  {journal} {The
  Journal of Chemical Physics}\ }\textbf {\bibinfo {volume} {72}},\ \bibinfo
  {pages} {2384--2393} (\bibinfo {year} {1980})}\BibitemShut {NoStop}%
\bibitem [{\citenamefont {Bennett}(1976)}]{bennett1976efficient}%
  \BibitemOpen
  \bibfield  {author} {\bibinfo {author} {\bibfnamefont {C.~H.}\ \bibnamefont
  {Bennett}},\ }\bibfield  {title} {\enquote {\bibinfo {title} {{Efficient
  estimation of free energy differences from Monte Carlo data}},}\ }\href
  {\doibase 10.1016/0021-9991(76)90078-4} {\bibfield  {journal} {\bibinfo
  {journal} {Journal of Computational Physics}\ }\textbf {\bibinfo {volume}
  {22}},\ \bibinfo {pages} {245--268} (\bibinfo {year} {1976})}\BibitemShut
  {NoStop}%
\bibitem [{\citenamefont {Voigt}\ \emph {et~al.}(2001)\citenamefont {Voigt},
  \citenamefont {Bienfait}, \citenamefont {Wang},\ and\ \citenamefont
  {Nicklaus}}]{voigt2001comparison}%
  \BibitemOpen
  \bibfield  {author} {\bibinfo {author} {\bibfnamefont {J.~H.}\ \bibnamefont
  {Voigt}}, \bibinfo {author} {\bibfnamefont {B.}~\bibnamefont {Bienfait}},
  \bibinfo {author} {\bibfnamefont {S.}~\bibnamefont {Wang}}, \ and\ \bibinfo
  {author} {\bibfnamefont {M.~C.}\ \bibnamefont {Nicklaus}},\ }\bibfield
  {title} {\enquote {\bibinfo {title} {Comparison of the nci open database with
  seven large chemical structural databases},}\ }\href@noop {} {\bibfield
  {journal} {\bibinfo  {journal} {Journal of chemical information and computer
  sciences}\ }\textbf {\bibinfo {volume} {41}},\ \bibinfo {pages} {702--712}
  (\bibinfo {year} {2001})}\BibitemShut {NoStop}%
\bibitem [{\citenamefont {Menichetti}, \citenamefont {Kremer},\ and\
  \citenamefont {Bereau}(2018)}]{Menichetti2018}%
  \BibitemOpen
  \bibfield  {author} {\bibinfo {author} {\bibfnamefont {R.}~\bibnamefont
  {Menichetti}}, \bibinfo {author} {\bibfnamefont {K.}~\bibnamefont {Kremer}},
  \ and\ \bibinfo {author} {\bibfnamefont {T.}~\bibnamefont {Bereau}},\
  }\bibfield  {title} {\enquote {\bibinfo {title} {{Efficient potential of mean
  force calculation from multiscale simulations: Solute insertion in a lipid
  membrane}},}\ }\href {\doibase 10.1016/J.BBRC.2017.08.095} {\bibfield
  {journal} {\bibinfo  {journal} {Biochemical and Biophysical Research
  Communications}\ }\textbf {\bibinfo {volume} {498}},\ \bibinfo {pages}
  {282--287} (\bibinfo {year} {2018})}\BibitemShut {NoStop}%
\bibitem [{\citenamefont {Kanekal}\ and\ \citenamefont
  {Bereau}(2019)}]{zenodo}%
  \BibitemOpen
  \bibfield  {author} {\bibinfo {author} {\bibfnamefont {K.~H.}\ \bibnamefont
  {Kanekal}}\ and\ \bibinfo {author} {\bibfnamefont {T.}~\bibnamefont
  {Bereau}},\ }\href {\doibase 10.5281/zenodo.3271766} {\enquote {\bibinfo
  {title} {{Resolution limit of data-driven coarse-grained models spanning
  chemical space}},}\ }\bibinfo {howpublished}
  {\url{http://doi.org/10.5281/zenodo.3271766}} (\bibinfo {year}
  {2019})\BibitemShut {NoStop}%
\end{thebibliography}%


\providecommand{\noopsort}[1]{}\providecommand{\singleletter}[1]{#1}%
%


\providecommand{\noopsort}[1]{}\providecommand{\singleletter}[1]{#1}%
\begin{thebibliography}{2}%
\makeatletter
\providecommand \@ifxundefined [1]{%
 \@ifx{#1\undefined}
}%
\providecommand \@ifnum [1]{%
 \ifnum #1\expandafter \@firstoftwo
 \else \expandafter \@secondoftwo
 \fi
}%
\providecommand \@ifx [1]{%
 \ifx #1\expandafter \@firstoftwo
 \else \expandafter \@secondoftwo
 \fi
}%
\providecommand \natexlab [1]{#1}%
\providecommand \enquote  [1]{``#1''}%
\providecommand \bibnamefont  [1]{#1}%
\providecommand \bibfnamefont [1]{#1}%
\providecommand \citenamefont [1]{#1}%
\providecommand \href@noop [0]{\@secondoftwo}%
\providecommand \href [0]{\begingroup \@sanitize@url \@href}%
\providecommand \@href[1]{\@@startlink{#1}\@@href}%
\providecommand \@@href[1]{\endgroup#1\@@endlink}%
\providecommand \@sanitize@url [0]{\catcode `\\12\catcode `\$12\catcode
  `\&12\catcode `\#12\catcode `\^12\catcode `\_12\catcode `\%12\relax}%
\providecommand \@@startlink[1]{}%
\providecommand \@@endlink[0]{}%
\providecommand \url  [0]{\begingroup\@sanitize@url \@url }%
\providecommand \@url [1]{\endgroup\@href {#1}{\urlprefix }}%
\providecommand \urlprefix  [0]{URL }%
\providecommand \Eprint [0]{\href }%
\providecommand \doibase [0]{http://dx.doi.org/}%
\providecommand \selectlanguage [0]{\@gobble}%
\providecommand \bibinfo  [0]{\@secondoftwo}%
\providecommand \bibfield  [0]{\@secondoftwo}%
\providecommand \translation [1]{[#1]}%
\providecommand \BibitemOpen [0]{}%
\providecommand \bibitemStop [0]{}%
\providecommand \bibitemNoStop [0]{.\EOS\space}%
\providecommand \EOS [0]{\spacefactor3000\relax}%
\providecommand \BibitemShut  [1]{\csname bibitem#1\endcsname}%
\let\auto@bib@innerbib\@empty
%</preamble>
\bibitem [{\citenamefont {Bereau}\ and\ \citenamefont
  {Kremer}(2015)}]{Bereau2015}%
  \BibitemOpen
  \bibfield  {author} {\bibinfo {author} {\bibfnamefont {T.}~\bibnamefont
  {Bereau}}\ and\ \bibinfo {author} {\bibfnamefont {K.}~\bibnamefont
  {Kremer}},\ }\href@noop {} {\bibfield  {journal} {\bibinfo  {journal}
  {Journal of chemical theory and computation}\ }\textbf {\bibinfo {volume}
  {11}},\ \bibinfo {pages} {2783} (\bibinfo {year} {2015})}\BibitemShut
  {NoStop}%
\bibitem [{\citenamefont {Marrink}\ \emph {et~al.}(2007)\citenamefont
  {Marrink}, \citenamefont {Risselada}, \citenamefont {Yefimov}, \citenamefont
  {Tieleman},\ and\ \citenamefont {Vries}}]{Marrink2007}%
  \BibitemOpen
  \bibfield  {author} {\bibinfo {author} {\bibfnamefont {S.~J.}\ \bibnamefont
  {Marrink}}, \bibinfo {author} {\bibfnamefont {H.~J.}\ \bibnamefont
  {Risselada}}, \bibinfo {author} {\bibfnamefont {S.}~\bibnamefont {Yefimov}},
  \bibinfo {author} {\bibfnamefont {D.~P.}\ \bibnamefont {Tieleman}}, \ and\
  \bibinfo {author} {\bibfnamefont {A.~H.~D.}\ \bibnamefont {Vries}},\ }\href
  {\doibase 10.1021/jp071097f} {\bibfield  {journal} {\bibinfo  {journal}
  {Journal of Physical Chemistry B}\ }\textbf {\bibinfo {volume} {111}},\
  \bibinfo {pages} {7812} (\bibinfo {year} {2007})}\BibitemShut {NoStop}%
\end{thebibliography}%

\end{document}

\title{Supporting Information to ``Resolution limit of data-driven coarse-grained models spanning chemical space''}

\author{Kiran H.~Kanekal}
\author{Kurt Kremer}
\author{Tristan Bereau}
\affiliation{Max Planck Institute for Polymer Research, 
  Ackermannweg 10, 55128 Mainz, Germany}

\date{\today}

\maketitle

\section{Introduction}
In this supplementary text, we report additional results referenced 
in the main paper. In Sec.~\ref{am_update} we detail the changes made to the previously published version of the \textsc{auto-martini} code. In Sec.~\ref{hist_text} we show the histograms comparing the $\dGwol$ distributions for each of the force fields studied in this text. In Sec.~\ref{gdb_dists} we provide statistics for each of the clusters used in the creation of the nine-bead-type force field.  In Sec.~\ref{ff_param} we demonstrate the result of interpolating across the Martini interaction matrix in order to parameterize our new force fields.  In Sec.~\ref{fg_analysis}, we provide lists of all the functional group pairs included in the calculation of the Bayesian likelihood and posterior distributions. We also include another analysis using this Bayesian approach for fragments that contain five heavy atoms. In addition, we have included text files containing force field parameters for each of the new force fields, the database of GDB compounds mapping to CG unimers and dimers for each force field, and trajectories for each of the simulations referenced in the main text in a zenodo repository which can be accessed via the following link: http://doi.org/10.5281/zenodo.3271766.

\section{Updates to \textsc{auto-martini}}
\label{am_update}

\begin{figure}[htbp]
  \begin{center}
    \includegraphics[width=\linewidth]{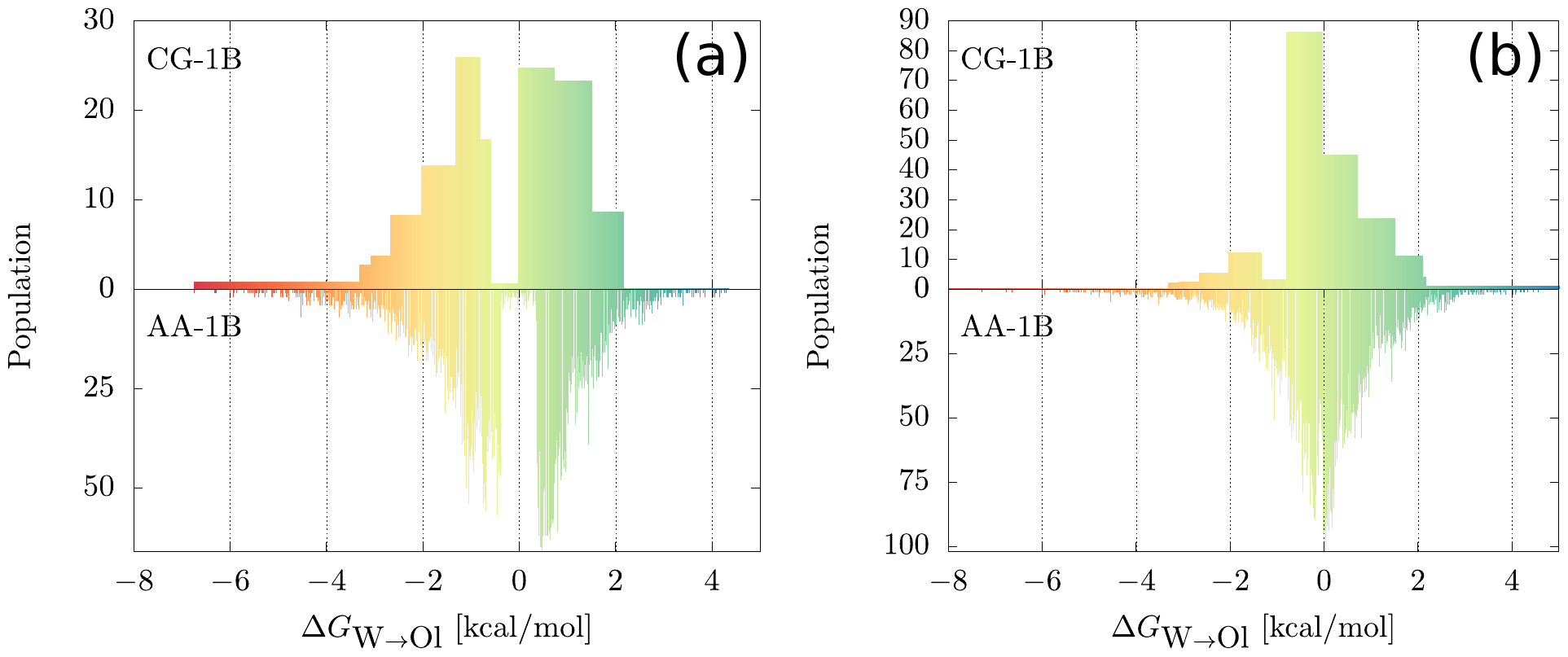}
    \caption{Comparison of the $\dGwol$ distributions for molecules mapping to a single Martini bead using the (a) originally published \textsc{auto-martini} code and (b) after increasing the lonely atom penalty and removing the additivity requirement for single beads. }
    \label{fig:master_vs_refactor}
  \end{center}
\end{figure}

Several changes were made to the \textsc{auto-martini} code in order to increase its accuracy when applied to a large and varied database such as the GDB. The ``lonely atom penalty''\cite{Bereau2015}, which weights the effect of leaving single heavy atoms outside the van der Waals radii of the Martini beads, was increased slightly from 0.20 to 0.28. Additionally, the ``additivity check'' was removed for molecules that map to single beads. This additivity check was designed to ensure that the voronoi decomposition of molecules into fragments and the subsequent selection of bead types for each fragment was sensible (the sum of the $\dGwol$ values for each bead should be within a cutoff value when compared to the $\dGwol$ of the entire molecule). This was enacted in order to resolve an issue in which molecules that were meant to be mapped to a single bead (e.g. Propanol) were unable to be successfully mapped using the code. The effect of these two changes on the distributions of $\dGwol$ is shown in Fig.~\ref{fig:master_vs_refactor}a and b. 

\begin{figure}[htbp]
  \begin{center}
    \includegraphics[width=\linewidth]{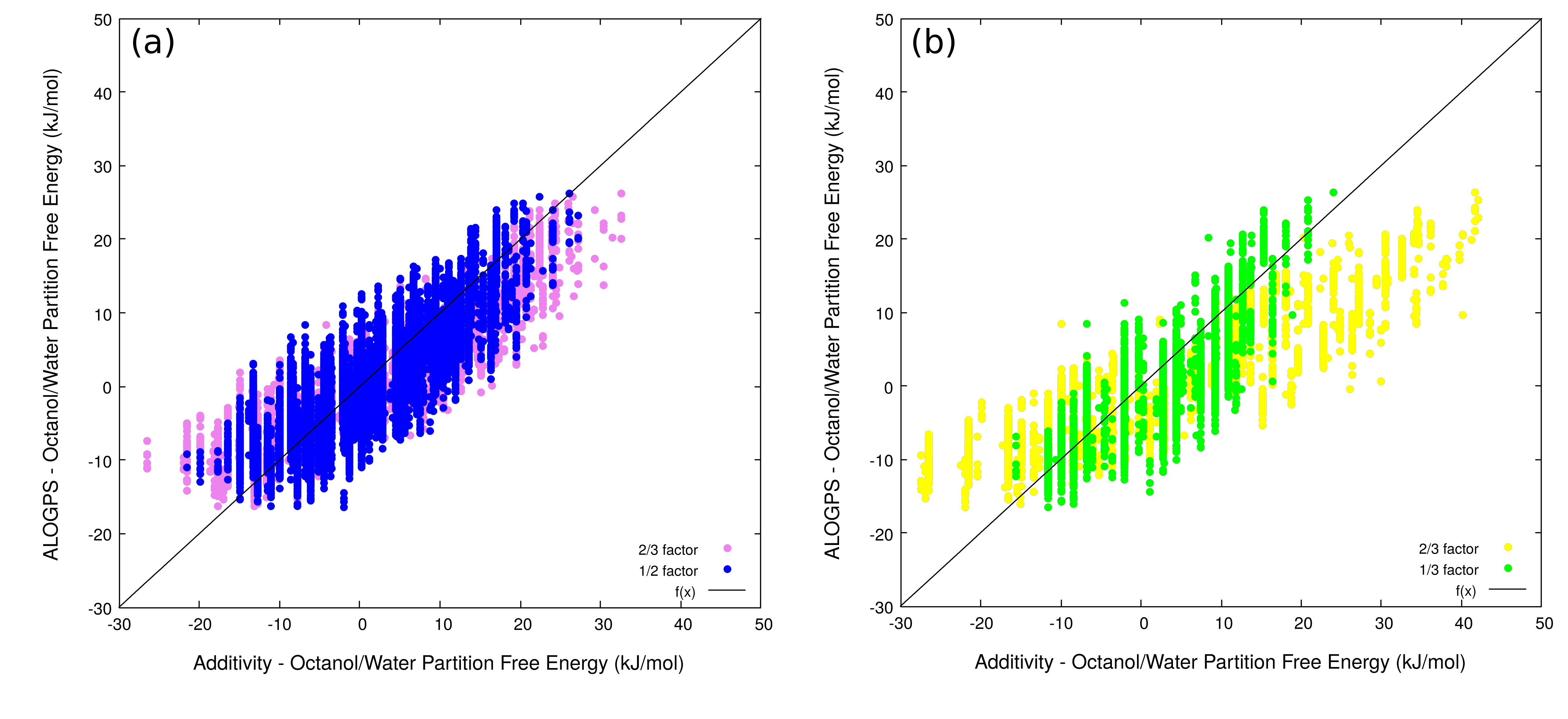}
    \caption{Correlation curves showing the agreement between the predicted partition free energy values from \textsc{ALOGPS} for ring molecules and the partition free energy of the coarse-grained Martini representation assigned by \textsc{auto-martini} for (a) five-membered and (b) six-membered ring-containing molecules.}
    \label{fig:ringfix}
  \end{center}
\end{figure}

The removal of the additivity check for molecules mapping to single beads caused the gap in the distribution in Fig.~\ref{fig:master_vs_refactor}a to no longer appear, meaning that several molecules that would normally map to a single bead were excluded because they failed the additivity check, which should not be applied for single beads. Note that there is a noticeable dip in the coarse-grained distribution of Fig.~\ref{fig:master_vs_refactor}. This corresponds to the N0 bead type, which is underpopulated when compared to the corresponding region in the atomistic distribution. We found that this was an artifact due to a cut-off value in the code that caused molecules to be mapped to a donor-acceptor type of bead even if their $\dGwol$ was closer to the N0 value. By reducing this cut-off value, we were able to obtain the distribution shown in Fig. 2a of the main text, and is also shown in Fig.~\ref{fig:Histcomb_SI}c. The final change has to do with the assignment of ring molecules. The standard approach for ring molecules was to use the entire set of atoms in the ring for each fragment and weight each bead's contribution by a scaling factor. For all ring molecules, this was previously set to 2/3 so as to reproduce the Martini parameterizations for benzene and cyclohexane.\cite{Bereau2015,Marrink2007} However, in order to optimize the mappings for the multitudes of ring-containing molecules in the GDB, we found that a factor of 1/2 for 5-membered rings and 1/3 for six-membered rings yielded much better agreement with respect to the \textsc{alogps} predictions for the ring molecules. The results are shown in Fig.~\ref{fig:ringfix} for both 5-membered and 6-membered rings. Changing these scaling factors resulted in an decrease in the MAE from 1.64 kcal/mol to 0.946 kcal/mol for 6-membered rings and a decrease from 0.893 kcal/mol to 0.807 kcal/mol for the 5-membered rings.

\section{Histograms}
\label{hist_text}

\begin{figure}[htbp]
  \begin{center}
    \includegraphics[width=0.9\linewidth]{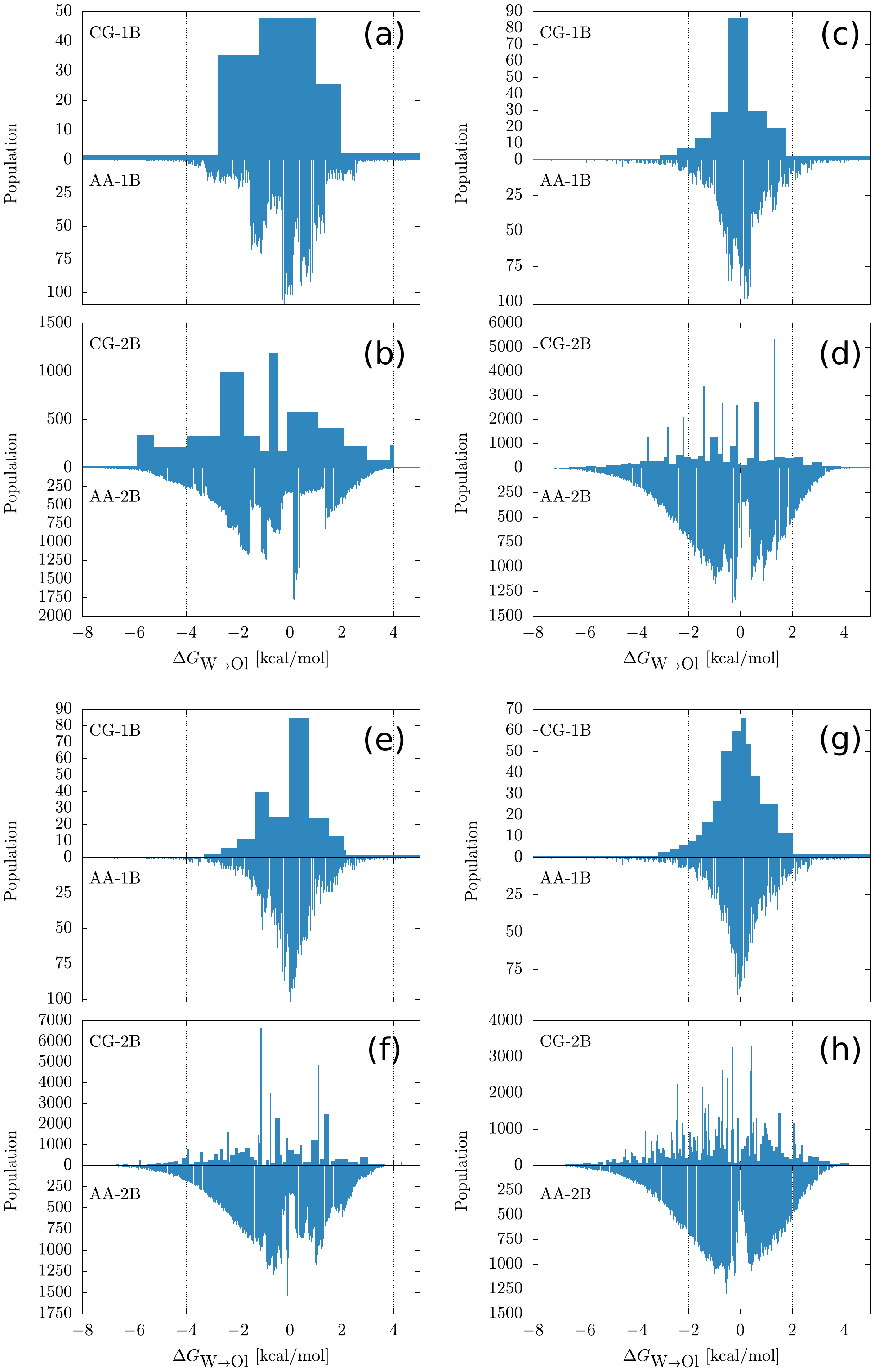}
    \caption{Histograms used to calculate JSD values shown in Fig. 2 of the main text. The unimer and dimer distributions are shown for the five-bead-type (a,b), nine-bead-type (c,d), Martini (e,f), and sixteen-bead-type (g,h) force fields. }
    \label{fig:Histcomb_SI}
  \end{center}
\end{figure}

In Fig.~\ref{fig:Histcomb_SI}, we show all of the histograms used to compute the JSDs shown in Fig. 2e of the main text. Note the significant differences in the $\dGwol$ distributions for the molecules mapping to dimers in the five-bead-type model (Fig.~\ref{fig:Histcomb_SI}b). The five-bead-type model is sampling a set of molecules from the GDB that is clearly different from those of the other models. While the other distributions contain populations ranging from $3.3 \cdot 10^5$ to $3.4 \cdot 10^5$, the 5-bead-type force field has only $3.0 \cdot 10^5$ molecules. Furthermore, even though the shapes of the distributions for the other three force fields are far more similar to each other than to the five-bead-type force field, the intersection of the sets of atomistic compounds mapping to each force field consists of ~$2.3 \cdot 10^5$ molecules. Including the set of molecules mapping to the 5-bead-type force field reduces this intersection to ~$1.8 \cdot 10^5$ molecules. This explains why the JSD value for the five-bead-type model is significantly lower than all of the others. 

\section{$\dGwol$ Distributions from the GDB}
\label{gdb_dists}
In the zenodo repository linked above, we provide plots of distributions of $\dGwol$ for molecules in the GDB containing up to eight heavy atoms. The distributions are constructed based on the number and type of heavy atom substitutions that exist in the molecules. For example, the file named ``GDB02to08\_HAstats\_fooo\_subs.pdf'' shows the $\dGwol$ distribution for all molecules containing one fluorine and three oxygen substitutions. Also included in the repository is a single file called ``GDB02to08\_HAstats.dat'' which contains the mean and standard deviation for each of the distributions provided, which were used to make Fig. 3 of the main text.

\section{Force field parameterization}
\label{ff_param}

\begin{figure}[htbp]
  \begin{center}
    \includegraphics[width=\linewidth]{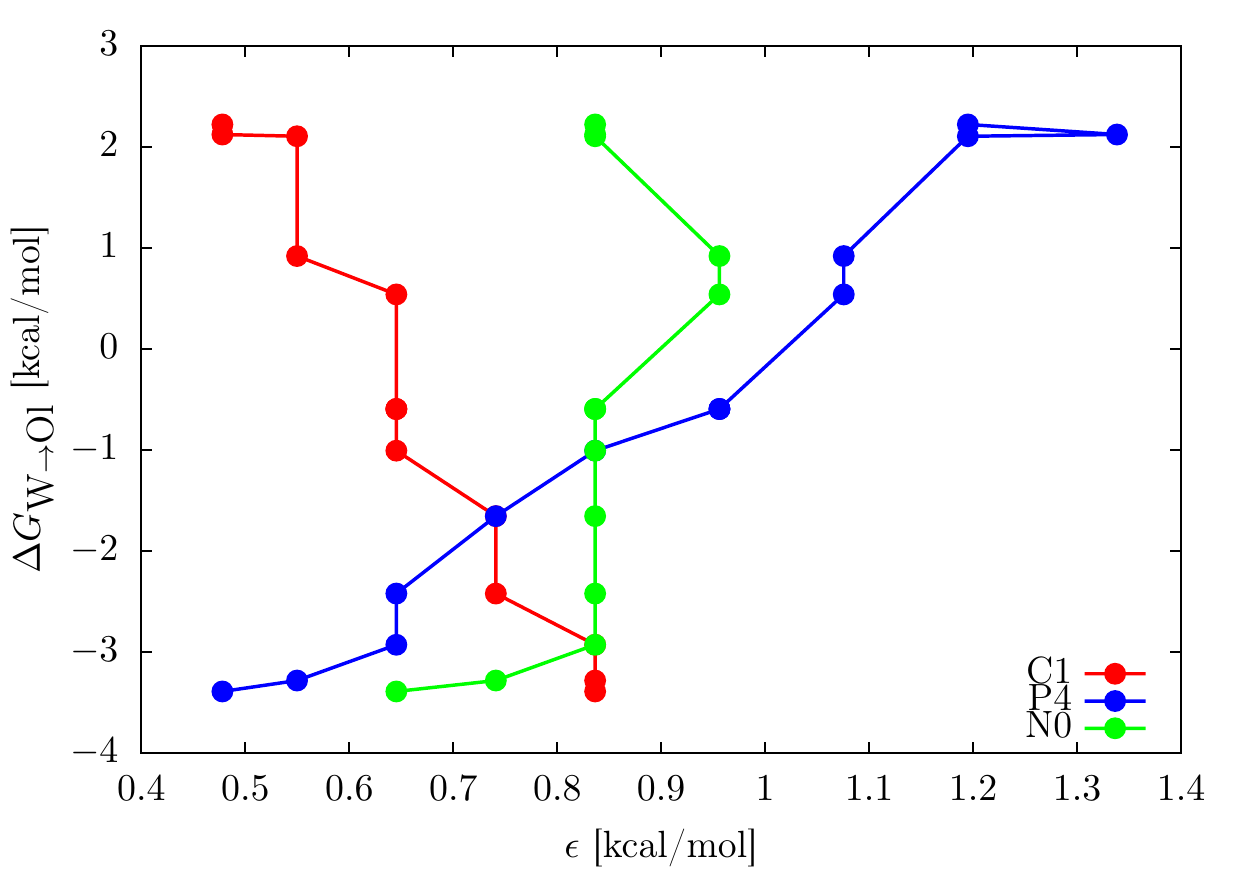}
    \caption{Relationship between the Lennard-Jones $\epsilon$ parameters for the Martini C1, P4, and N0 bead types and the $\dGwol$ values for every Martini bead type.}
    \label{fig:eps_vs_dGwol}
  \end{center}
\end{figure}

\begin{figure}[htbp]
  \begin{center}
    \includegraphics[width=\linewidth]{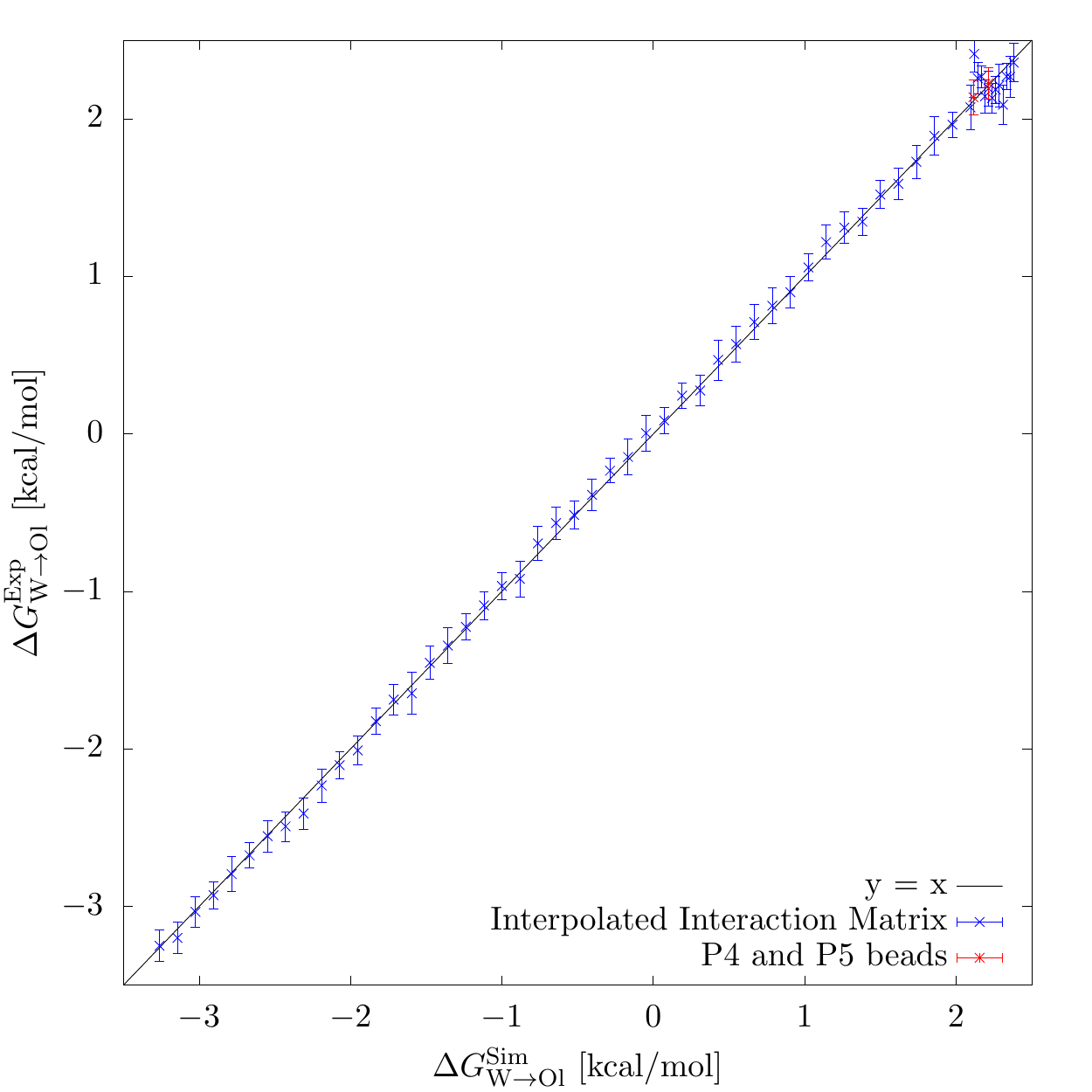}
    \caption{Calibration curve showing desired $\dGwol$ values on the horizontal axis and $\dGwol$ values obtained from simulations using beads parameterized by interpolating across the Martini interaction matrix. }
    \label{fig:interp_curve}
  \end{center}
\end{figure}

Fig.~\ref{fig:eps_vs_dGwol} shows the relationship between the Lennard-Jones $\epsilon$ parameter (related to the depth of the attractive well) for one of three given Martini bead types and the $\dGwol$ for all Martini bead types.\cite{Marrink2007} It is evident that there is no clear underlying functional form that can be applied to all Martini bead types. While there are localized regions that can be easily fit to lines, there are sharp discontinuities for each of the bead types at the boundaries of these localized regions. Therefore, linear interpolation is used to preserve these discontinuities in the new models, using the desired $\dGwol$ as the target. The results of this interpolation are shown in Fig.~\ref{fig:interp_curve}. To construct this plot, we parameterized a new bead type for a series of $\dGwol$ values evenly spaced along the range of $\dGwol$ covered by Martini and ran simulations to calculate the $\dGwol$ for each using the methods described in the main text. The results clearly show that the interpolation was successful for recovering the desired $\dGwol$. There are some slightly larger deviations close for $\dGwol$ closer to the P4 and P5 beads. This is probably due to the fact that, despite having a more attractive interaction with Martini water, the P5 bead has a slightly more positive $\dGwol$ than the P4 bead type.\cite{Bereau2015}

Using this interpolation method, we parameterized three CG force fields. For the donor and acceptor types, we assigned the bead type which had the $\dGwol$ closest to 0.0 as the bead type corresponding to molecules containing both donor and acceptor (da) groups. We then followed the example set in the Martini interaction matrix.\cite{Marrink2007} The donor-only and acceptor-only bead type were assigned the same parameters as the da bead type but with a decrease in the $\epsilon$ value of 0.5 kJ/mol (making the interaction slightly more repulsive) when interacting with like bead types. For each force field, the bead types and corresponding $\dGwol$ values are given below.

\FloatBarrier

\begin{table}[!htb]
\begin{center}
\begin{tabular}{| c | c | c |}
\hline
Bead Type Name & Polar/Nonpolar/Apolar, Donor/Acceptor  & $\dGwol$ [kcal/mol]\\
\hline
T1 & Polar & 2.05 \\
T2 & Polar & 1.91 \\
T3 & Nonpolar Donor+Acceptor & 0.098 \\
T3d & Nonpolar Donor & 0.098 \\
T3a & Nonpolar Acceptor & 0.098 \\
T4 & Apolar & -2.46 \\
T5 & Apolar & -3.13 \\
\hline
\end{tabular}
\caption{Names, characteristics, and $/dGwol$ values for each bead type in the five-bead-type force field.
}
\label{table:5TTYPES}
\end{center}
\end{table}

\begin{table}[!htb]
\begin{center}
\begin{tabular}{| c | c | c |}
\hline
Bead Type Name & Polar/Nonpolar/Apolar, Donor/Acceptor  & $\dGwol$ [kcal/mol]\\
\hline
T1 & Polar & 2.14 \\
T2 & Polar & 1.39 \\
T3 & Polar & 0.672 \\
T4 & Nonpolar Donor+Acceptor & -0.074 \\
T4d & Nonpolar Donor & -0.074 \\
T4a & Nonpolar Acceptor & -0.074 \\
T5 & Nonpolar & -0.899 \\
T6 & Apolar & -1.36 \\
T7 & Apolar & -2.17 \\
T8 & Apolar & -2.76 \\
T9 & Apolar & -3.51 \\
\hline
\end{tabular}
\caption{Names, characteristics, and $/dGwol$ values for each bead type in the nine-bead-type force field.
}
\label{table:9TTYPES}
\end{center}
\end{table}

\begin{table}[!htb]
\begin{center}
\begin{tabular}{| c | c | c |}
\hline
Bead Type Name & Polar/Nonpolar/Apolar, Donor/Acceptor  & $\dGwol$ [kcal/mol]\\
\hline
P4 & Polar & 2.22 \\
P5 & Polar & 2.12 \\
P3 & Polar & 2.11\\
P2 & Polar & 0.92 \\
P1 & Polar & 0.54 \\
Nda & Nonpolar Donor+Acceptor  & -0.595 \\
Nd & Nonpolar Donor & -0.595 \\
Na & Nonpolar Acceptor & -0.595 \\
N0 & Nonpolar & -1.00 \\
C5 & Apolar & -1.66 \\
C4 & Apolar & -2.42 \\
C3 & Apolar & -2.93 \\
C2 & Apolar & -3.28 \\
C1 & Apolar & -3.39 \\
\hline
\end{tabular}
\caption{Names, characteristics, and $/dGwol$ values for each neutral bead type in the Martini force field.
}
\label{table:Martini}
\end{center}
\end{table}

\begin{table}[!htb]
\begin{center}
\begin{tabular}{| c | c | c |}
\hline
Bead Type Name & Polar/Nonpolar/Apolar, Donor/Acceptor  & $\dGwol$ [kcal/mol]\\
\hline
T1 & Polar & 2.18 \\
T2 & Polar & 1.85 \\
T3 & Polar & 1.03\\
T4 & Polar & 0.507 \\
T5 & Polar & 0.335 \\
T6 & Nonpolar & 0.126 \\
T7 & Nonpolar Donor+Acceptor  & -0.061 \\
T7d & Nonpolar Donor & -0.061 \\
T7a & Nonpolar Acceptor & -0.061 \\
T8 & Nonpolar & -0.627 \\
T9 & Apolar & -0.838 \\
T10 & Apolar & -1.33 \\
T11 & Apolar & -1.62 \\
T12 & Apolar & -1.82 \\
T13 & Apolar & -2.20 \\
T14 & Apolar & -2.62 \\
T15 & Apolar & -2.81 \\
T16 & Apolar & -3.60 \\
\hline
\end{tabular}
\caption{Names, characteristics, and $/dGwol$ values for each bead type in the sixteen-bead-type force field.
}
\label{table:16TTYPES}
\end{center}
\end{table}

\section{Functional group analysis}
\label{fg_analysis}

\begin{figure*}[htbp]
  \begin{center}
    \includegraphics[height= 0.9\textheight]{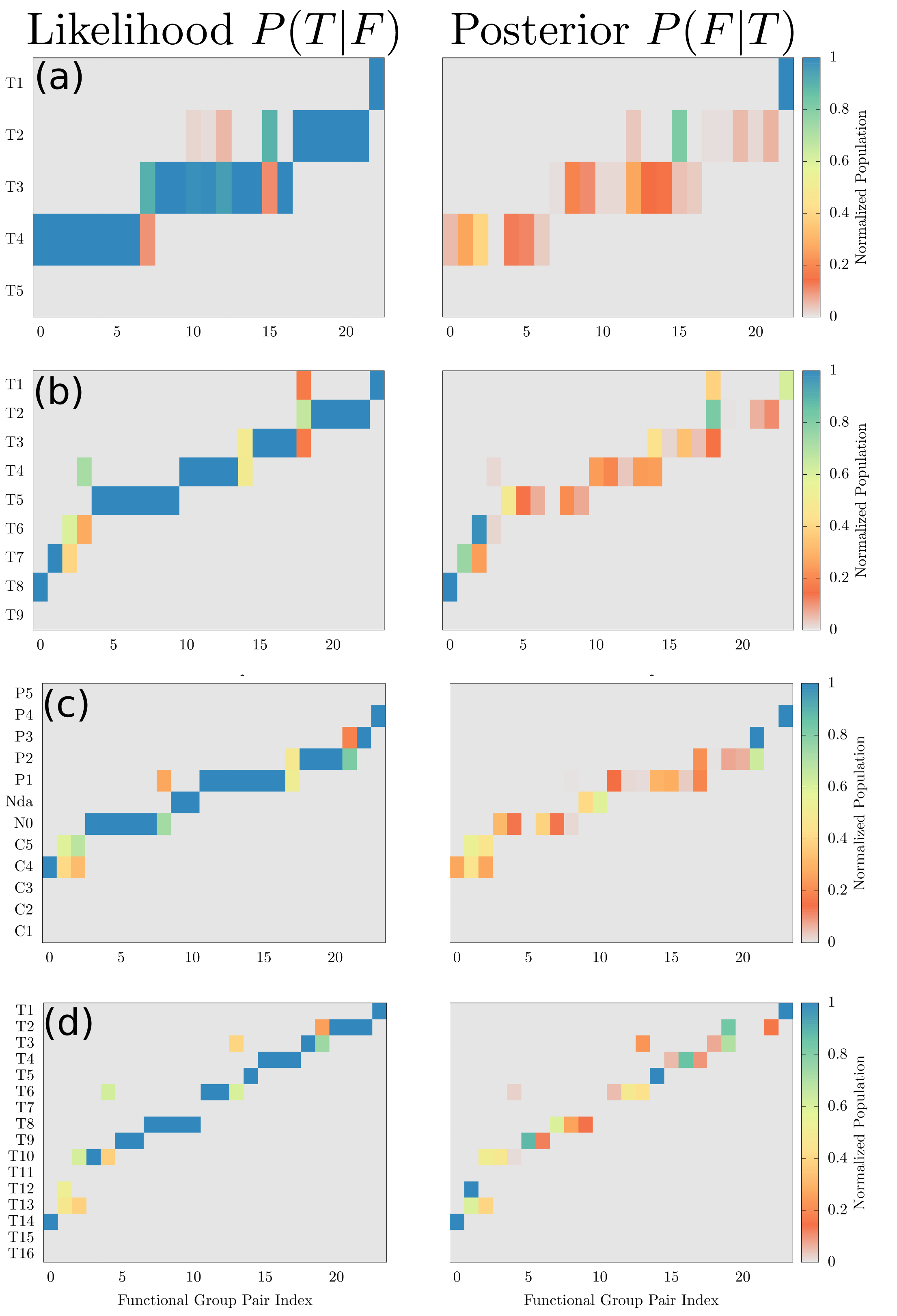}
    \caption{Heat maps portraying the degeneracy of specific pairs of functional groups for a given bead type for force fields containing five (a), nine (b), twelve (c), or sixteen (d) bead types. The horizontal axes denote specific functional group pairs that exist in a chemical fragment with four heavy atoms only. The color corresponds to either the Bayesian likelihood (left side) or posterior (right side) probabilities.}
    \label{fig:4HA_FGPairs}
  \end{center}
\end{figure*}

Fig.~\ref{fig:4HA_FGPairs} shows the likelihood and posterior values calculated for fragments containing only four heavy atoms and two functional groups as specified by \textsc{checkmol}. The total number of bead types of each force field is not reflected in these heat maps, with the most apolar bead types missing. This is because all of the fragments that map to these bead types consist of saturated hydrocarbons or single alkene/alkyne substitutions only, and thus are not detected as having a functional group pair by \textsc{checkmol}. Furthermore, there are no values calculated for the T7 beads in the sixteen-bead-type force field because there were no donor/acceptor/donor+acceptor fragments that also had two functional groups within the narrow range of $\dGwol$ covered by the T7 bead types. Similar reasoning can also be applied to explain the lack of values for the T11 bead type in the same force field.

Over the course of this work, certain idiosyncrasies were discovered when using \textsc{checkmol}. One such issue was the fact that the code tended to double-count some functional groups. For example, fragments with only a single fluorine substitution were counted as both a ``halogen derivative'' and as a ``alkyl fluoride''. This was only observed for the aforementioned fluorine substitutions as well as for dialkyl ethers. Other examples were also found for which the software could not correctly identify the functional groups contained in the fragment. This is probably due to the fact that \textsc{checkmol} was not tested on some of the less common chemistries encountered in the GDB. The most egregious example of this was found for fragments containing the smiles string ``NC=N'' which were incorrectly labeled as a carboxylic acid derivatives by \textsc{checkmol}. For this reason, we did not explicitly label the horizontal axes with their corresponding chemistries in Fig.~\ref{fig:4HA_FGPairs} and Fig. 4 of the main text. For full transparency, we have included the smiles string for each unique fragment used in the Bayesian analysis for both four-heavy-atom and five-heavy-atom fragments as well as the corresponding values for $P(F)$, $P(T)$, $P(T|F)$, and $P(F|T)$ in the zenodo repository. While the functional group labels given by \textsc{checkmol} are incorrect in a few cases, the overall trends reported in this work are unaffected: namely, that increasing the number of bead types also provides increased values of the Bayesian posterior probabilities for back-mapping specific chemistries.

\bibliography{biblio}